\RequirePackage{fix-cm}
\documentclass[twocolumn]{svjour3}          
\smartqed  

\usepackage{graphicx}
  \usepackage{dcolumn}
  \usepackage{bm}
  \usepackage{ulem}
  \usepackage[utf8]{inputenc}
  \usepackage[english]{babel}
  \graphicspath{{figs/}{figs:}{\figs}}
  \usepackage[usenames,dvipsnames]{color}
  \usepackage[]{ulem}
  \usepackage[colorlinks=true,
  linkcolor=blue,
  urlcolor=blue,
  citecolor=blue,
  breaklinks=true]{hyperref}
  \newcommand{\comment}[1]{\textcolor{red}{#1}}
  \renewcommand{\comment}[1]{\relax}
  \newcommand{\todelete}[1]{\textcolor{green}{\sout{#1}}}
  \renewcommand{\todelete}[1]{\relax}

\usepackage{mathtools}
\usepackage{multirow}
\usepackage{amssymb}
\usepackage{amsmath}
\usepackage{subfigure}

\usepackage{float}
\begin{document}

\title{Floquet analysis on a viscous cylindrical fluid surface subject to a time-periodic radial acceleration
}


\author{Dilip Kumar Maity
}


\institute{Dilip Kumar Maity, Indian Institute of Technology Kharagpur, Kharagpur-721302, India\\
             \email{dilmaity@gmail.com} 
}

\date{Received: date / Accepted: date}

\maketitle

\begin{abstract}
	Parametrically excited standing waves are observed on a cylindrical fluid filament. These are the cylindrical analog of Faraday instability in a flat surface or spherical droplet. Using the Floquet technique, linear stability analysis has been investigated on a viscous cylindrical fluid surface, which is subjected to a time-periodic radial acceleration. Viscosity has a significant impact on the critical forcing amplitude as well as the dispersion relation of the non-axis symmetric patterns. The effect of viscosity on onset parameters of the pattern with azimuthal wavenumber, $m=1$, has shown different dependency from $m>1$. The effect of viscosity increases with an increasing $m$ is also observed.   
\end{abstract}

\section{\label{sec:level1}Introduction}
Interfacial instability occurs between the surfaces of two different fluids due to the application of a time-periodic vertical acceleration, has been known as Faraday Instability. This type of parametrically excited standing wave pattern was first observed by Faraday{~\cite{Faraday_1831}} in 1831 [see for a review{~\cite{Miles_Henderson_1990}}]. The excited surface waves have a frequency either half{~\cite{Faraday_1831}} or equal to{~\cite{Matthiessen_1868,Kumar_1996}} that of the forcing frequency. Experiments{~\cite{Faraday_1831,Tufillaro_1989,Ciliberto_1991} with low viscous fluid have shown a square pattern at the near onset. Benjamin and Ursell{~\cite{Benjamin_1954}} first came with a theory where they converted the linear non-viscous dynamical equations to a Mathieu equation. According to their theory, the fluid surface is unstable inside the tongue-like boundary, which is plotted in the parameter space of the axial wavenumber $k$ and the forcing amplitude $a$. Later experiments with viscous fluid showed stripes{~\cite{fauve}}, regular tri-angular patterns{~\cite{Muller}}, competing hexagons and triangles pattern{~\cite{Bajaj}}. These observations could not be explained by the theory of Benjamin and Ursell{~\cite{Benjamin_1954}}. Kumar{~\cite{Kumar_1996}}, and Kumar and Tuckerman{~\cite{Kumar_1994}} converted linear viscous Navier-Stokes equations to a finding of eigenvalue problem using Floquet theory. This fundamental viscous theory{~\cite{Kumar_1996,Kumar_1994}} helps to understand the mechanisms behind pattern selection on the fluid surface.
	
	On the other hand, instability on the curved fluid surface{~\cite{Kelvin_1863,Rayleigh,Lamb_1932,Reid,Chandrasekhar_1961,Miller_Scriven,Plateau,Bohr,Popinet,Cervone,Dasgupta,Shen,Hill,Noblin,Brunet,Adou_Tuckerman_2016,Li,Dilip,Dasgupta_2,Wang,Futterer,Falcon}} also has a great interest in the scientific community. These have a wide application in measuring surface tension, studying pattern formation, micro-fluidic devices, controlling jet breakup, fluid atomization, coating, and drug delivery, etc. A sphere and a cylinder are the two basic elements in the group of curved geometry. A spherical liquid drop and a cylindrical jet or filament are examples of two basic states which are formed due to surface tension.
	
	Kelvin{~\cite{Kelvin_1863}}, Rayleigh{~\cite{Rayleigh}}, and Lamb{~\cite{Lamb_1932}} derived natural frequency of the spherical drop when it is slightly perturbed from its equilibrium shape. Viscous correction of this problem was accomplished by Reid{~\cite{Reid}}, Chandrasekhar{~\cite{Chandrasekhar_1961}} and Miller and Scriven{~\cite{Miller_Scriven}}.
	When a liquid jet is falling under gravity, it breaks up into a smaller packet of equal volume, but with less surface area is called Rayleigh-Plateau (RP) instability. Plateau{~\cite{Plateau}} and Rayleigh{~\cite{Rayleigh}} derived dispersion relation of the liquid jet. Plateau{~\cite{Plateau}} remarked that a liquid jet was stable for all purely non-axis symmetric deformations and was unstable for axis-symmetric varicose deformations with wavelengths exceeding the circumference of the cylinder. Later viscous-extension of the dispersion relation of the axis-symmetric mode of RP instability was executed by Rayleigh{~\cite{Rayleigh}}, Bohr{~\cite{Bohr}}, and Chandrasekhar{~\cite{Chandrasekhar_1961}}. These relations were verified by using Direct numerical simulation (DNS){~\cite{Popinet,Cervone,Dasgupta}}. 
	
	When acoustically{~\cite{Shen}} or magnetically{~\cite{Hill}} levitated the spherical liquid drop is subjected to a time-periodic modulation, or the drops or puddles{~\cite{Noblin,Brunet}} are weakly pinned to a time-periodic oscillating surface, star-like patterns are excited. These observations are emphatically explained by Adou and Tuckerman{~\cite{Adou_Tuckerman_2016}}, considering the Faraday instability on the surface of the spherical drops subject to a time-periodic radial vibration. They did Floquet analysis, which is the spherical analog of the work of Kumar{~\cite{Kumar_1996}}, and Kumar and Tuckerman{~\cite{Kumar_1994}}. Using the Floquet analysis, Li et al. {~\cite{Li}} showed the effect of outside inviscid medium on the Faraday instability over the surface of the spherical drop.  Our recent experiment{~\cite{Dilip}} shows a half-cylindrical water surface, whose radius is much less than its length, where different types of non-axis symmetric patterns are parametrically excited due to the application of vertical vibration. Also, another recent study{~\cite{Dasgupta_2}} shows different non-axis-symmetric patterns, which are excited on the interface of a cylindrical filament subject to a time-periodic radial acceleration. However, in these studies{~\cite{Dilip,Dasgupta_2}}, the stable modes of a cylindrical jet under RP instability was unstable due to application of the time-periodic acceleration. 
	
	In this article, linear stability analysis is carried out on a viscous cylindrical fluid surface subject to a time-periodic radial acceleration. As the main focus of the article is on the surface tension driven instability, i.e., capillary instability, the effect of gravity is neglected throughout my study. In laboratory it is very difficult to produce radially directed acceleration. It can be produce in gravity free region{~\cite{Adou_Tuckerman_2016,Dasgupta_2,Wang,Futterer,Falcon}}. It is a cylindrical extension of the work done by Kumar{~\cite{Kumar_1996}}, and Kumar and Tuckerman{~\cite{Kumar_1994}} in plane geometry and Adou and Tuckerman{~\cite{Adou_Tuckerman_2016}}, and Li et al. {~\cite{Li}} in spherical drop. The effect of viscosity on the instability onsets of the non-axis symmetric patterns is presented here in detail. The paper is presented as follows in Sec.II, the mathematical formulations, including the governing hydrodynamical equations and linearization of the equations are presented. In Sec.III, solutions are done using the Floquet method for the ideal and viscous fluids. The results are also discussed in this section in detail, finally, in Sec.IV, conclusions are made over the whole manuscript.
	\section{Mathematical formulations}
	
	\subsection{Governing hydrodynamic equations of motion}
	Consider an incompressible viscous fluid element which is confined by an infinite cylindrical free surface of radius $R$, is subjected to a uniform radial sinusoidal acceleration. A cylindrical coordinate system is chosen with the $z$ axis coincident with the axis of the cylinder. The angle $\theta$ is measured from the $+x$ axis in counter-clockwise direction. In Fig.~\ref{fig1}, a schematic diagram of the considered system is shown. The motion of fluid element is governed by Navies-Stokes equations.
	
	\begin{equation}\label{nonlinear_navier_stokes}
	\rho\left[ \frac{\partial{\boldsymbol U}}{\partial{t}}
	+({\boldsymbol U}\cdot \boldsymbol {\nabla}){\boldsymbol U}\right]
	=-\boldsymbol{\nabla}P+\mu {\nabla}^2 
	{\boldsymbol U}
	+\rho \textbf G(t),
	\end{equation}
	\begin{equation} \label{continuity}
	\boldsymbol{\nabla \cdot U} = 0,
	\end{equation}
	where $\boldsymbol U$, $P$, $\rho$ and $\mu$ represents velocity, pressure, density and dynamical viscosity of the fluid, respectively. Throughout the manuscript vector quantities are presented using bold representation of it. $\boldsymbol{\hat{r}}$ represents unit vector along the radial direction in cylindrical coordinate system ($r,\theta,z$) . $\textbf G(t)$ is external sinusoidal radial acceleration given by 
	\begin{equation} \label{periodic_forcing}
	\textbf G(t)= -a\cos (\omega t) \hat{\boldsymbol{r}}=G(t) \hat{\boldsymbol{r}}.
	\end{equation}
	Where $\omega=2 \pi f$; $f$, $a$ are external driving frequency and amplitude, respectively. As for viscous fluid $a>>g$, the effect of gravitational acceleration, $g$, is neglected in the study.  
	
	\begin{figure*}[]
		\begin{center}
			\includegraphics[height=!, width=14 cm]{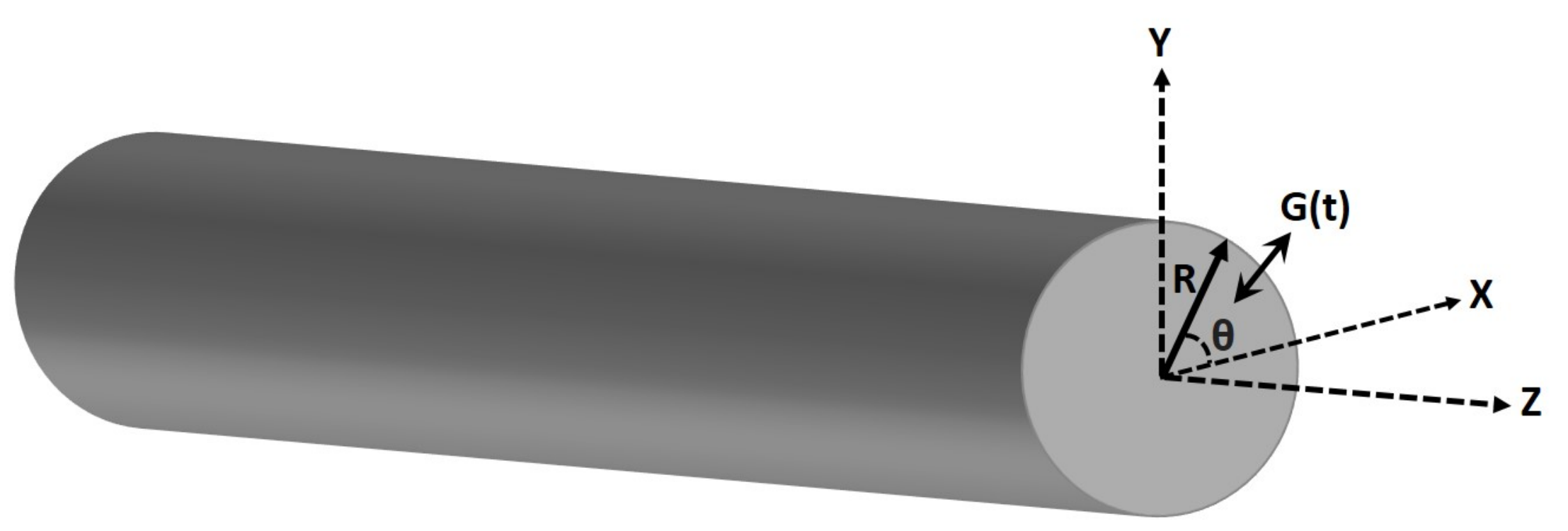}
			\caption{{Schematic diagram of a static cylindrical fluid surface subject to a time periodic radial acceleration.}
			} \label{fig1}
		\end{center}
	\end{figure*}
	
	When the surface of fluid is deformed due to external acceleration, then fluid interface is located at 
	\begin{equation} \label{interface_deformation}
	r=R+\eta (\theta ,z,t),
	\end{equation}
	where $\eta$ is the surface deformation.
	
	The surface of fluid is defined by 
	\begin{equation} \label{surface_equation}
	F(r,\theta,z,t)=r-[R+\eta (\theta ,z,t)]=0.
	\end{equation}
	
	The kinematic boundary condition at the fluid surface is given by 
	\begin{equation} \label{kinematic_boundary}
	\frac{\partial{F}}{\partial{t}}
	+({\boldsymbol U}\cdot \boldsymbol {\nabla}){F}
	=0.
	\end{equation}
	
	Using Eq.~(\ref{surface_equation}), kinematics boundary condition reads as 
	\begin{equation} \label{kinematic_boundary_condition_nonlinear}
	\frac{\partial{\eta}}{\partial{t}}
	+({\boldsymbol U} \cdot \boldsymbol {\nabla}){\eta}
	={U_r}{\mid}_{r=R+\eta}.
	\end{equation}
	
	Interfacial stress balance condition is defined by 
	\begin{equation} \label{interface_stress_balance}
	{\boldsymbol n}\cdot {\boldsymbol {\hat{\Pi}}}-{\boldsymbol n}\cdot {\boldsymbol \Pi}=\sigma {\boldsymbol n} ({\boldsymbol \nabla}\cdot {\boldsymbol n})-{\boldsymbol \nabla} {\sigma},
	\end{equation}
	
	where $\sigma$ is the surface tension coefficient of the fluid, and $\boldsymbol n$ represents the unit outward normal to the surface. ${\boldsymbol \Pi}$ and ${\boldsymbol{\hat{\Pi}}}$ represent stress tensor of the working fluid and the outside medium, respectively.
	
	Stress tensor of the fluid element can be written as
	\begin{equation} \label{stress_tensor}
	{\boldsymbol n}\cdot{\boldsymbol \Pi}=-P\boldsymbol I_d+\mu [\boldsymbol \nabla \boldsymbol U+(\boldsymbol \nabla \boldsymbol U)^T],
	\end{equation}
	where $\boldsymbol I_d$ represent unit vectors.
	
	Neglecting the effect of the outside medium and considering uniform surface tension of the fluid, normal stress balance condition can be expressed as
	\begin{equation} \label{normal_stress_balance}
	-{\boldsymbol n}\cdot {\boldsymbol \Pi} \cdot {\boldsymbol n}=\sigma \boldsymbol{\nabla} \cdot {\boldsymbol n} = 
	\sigma\left[\frac{1}{R_1}+\frac{1}{R_2}\right],  
	\end{equation}
	where $R_1$ and $R_2$  are the principle radii of curvature at a given point of the surface.
	
	Tangential stress balance conditions can be written as 
	\begin{equation} \label{tangential_stress_balance}
	{\boldsymbol n}\cdot {\boldsymbol \Pi} \cdot {\boldsymbol t}=0,
	\end{equation}
	where  ${\boldsymbol t}$ represents both tangent vector along $\theta$ and $z$ direction.

	\subsection{Linearizing the governing equations}
	The velocity and the pressure  can be decomposed with respect to unperturbed state ($\bar {\boldsymbol U}$, $\bar { P}$), respectively as
	\begin{subequations}\label{decompose}
		\begin{align}
		\boldsymbol U & = \bar {\boldsymbol U} +\boldsymbol u, \\
		P & = \bar { P} + p,
		\end{align}
	\end{subequations}
	where $u$ and $p$ are perturbed velocity and pressure, respectively.

	So, linearization of governing equations (\ref{nonlinear_navier_stokes}) and (\ref{continuity}) lead to
	\begin{equation}\label{linear_navier_stokes}
	\rho \frac{\partial{\boldsymbol u}}{\partial{t}}
	=-\boldsymbol{\nabla}p+\mu {\nabla}^2 
	{\boldsymbol u},
	\end{equation}
	\begin{equation} \label{continuity_2}
	\boldsymbol{\nabla \cdot u} = 0.
	\end{equation}
	
	Linear stability analysis can be proceeded by assuming $\eta<<R$.
	Considering motionless fluid element ($\bar {\boldsymbol U}=0$) before onset, kinematic boundary condition, Eq.~(\ref{kinematic_boundary_condition_nonlinear}) can be linearized as
	\begin{equation} \label{linearised_kinematic}
	\frac{\partial{\eta}}{\partial{t}}
	={u_r}{\mid}_{r=R}.
	\end{equation}
	
	The component of the stress tensor in the cylindrical fluid surface can be written as 
	
	\begin{subequations}\label{component_stress_balance}
		\begin{align}
		\Pi_{r\theta} & =\mu\left( \frac{\partial{u_\theta}}{\partial{r}}-\frac{u_\theta}{r}+\frac{1}{r} \frac{\partial{u_r}}{\partial{\theta}}\right), \\
		\Pi_{rz} & =\mu\left( \frac{\partial{u_r}}{\partial{z}}+\frac{\partial{u_z}}{\partial{r}}\right), \\ 
		\Pi_{rr} & = 2\mu \frac{\partial{u_r}}{\partial{r}}.  
		\end{align}
	\end{subequations}
	
	Normal stress balance condition, described in Eq.~ (\ref{normal_stress_balance}), can be rewritten as 
	
	\begin{equation} \label{normal_stress_balance_linear}
	-[(-P\boldsymbol I_d+\mu [\boldsymbol \nabla \boldsymbol U+(\boldsymbol \nabla \boldsymbol U)^T]) \cdot {\boldsymbol n}]_{r=R+\eta} 
	= \sigma\left[\frac{1}{R_1}+\frac{1}{R_2}\right].  
	\end{equation}
	
	Linearization~\cite{pressure_gradient} of left hand side of Eq.~(\ref{normal_stress_balance_linear}), leads
	
	\begin{equation} \label{left_normal_stress_balance_linear}
	\begin{multlined}
	(\bar P+p){\mid}_{r=R+\eta}-2\mu{\frac{\partial{u_r}}{\partial{r}}}{\mid}_{r=R+\eta} 
	=\sigma/R+\rho G(t)\eta\\
	+p{\mid}_{r=R}-2\mu{\frac{\partial{u_r}}{\partial{r}}}{\mid}_{r=R}.
	\end{multlined}
	\end{equation}
	
	Whereas, right hand side of Eq.~(\ref{normal_stress_balance_linear}) can be linearized~\cite{Lamb_1932} as 
	
	\begin{equation} \label{right_normal_stress_balance_linear}
	\sigma\left[\frac{1}{R_1}+\frac{1}{R_2}\right]_{r=R+\eta}
	=\sigma/R-{\sigma}\left(\frac{1}{R^2}
	+\frac{1}{R^2}\frac{\partial^2{}}{\partial{\theta^2}}
	+\frac{\partial^2{}}{\partial{z^2}}\right)\eta.
	\end{equation}
	
	So, Eq.~(\ref{normal_stress_balance_linear}) can be written as  
	\begin{equation} \label{pressure_surface_1}
	p{\mid}_{r=R}=2\mu{\frac{\partial{u_r}}{\partial{r}}}{\mid}_{r=R}-\rho G(t)\eta-{\sigma}\left(\frac{1}{R^2}
	+\frac{1}{R^2}\frac{\partial^2{}}{\partial{\theta^2}}
	+\frac{\partial^2{}}{\partial{z^2}}\right)\eta.
	\end{equation}
	
	Tangential stress component given in Eq.~(\ref{tangential_stress_balance}) must vanish at the free surface, with linearization leading to 
	\begin{equation} \label{linear_tangential_stress_balance}
	\Pi_{r\theta}{\mid}_{r=R}=\Pi_{rz}{\mid}_{r=R}=0.
	\end{equation}
	
	Horizontal divergence of ($\Pi_{r\theta} \hat{\boldsymbol{\theta}}+\Pi_{rz} \hat{\boldsymbol{z}}$) is
	\begin{equation}\label{horizontal_divergence_tengential_stress_balance}
	\begin{split}
	0 & =[\boldsymbol \nabla_H \cdot (\Pi_{r\theta}  {\hat{\boldsymbol{\theta}}}+\Pi_{rz} \hat{\boldsymbol{z}})]_{r=R} \\
	& =\mu\left[\frac{1}{r}\frac{\partial{\Pi_{r\theta}}}{\partial{\theta}}+\frac{\partial{\Pi_{rz}}}{\partial{z}}\right]_{r=R} \\ 
	& = \mu\left[\left(\frac{\partial^2{}}{\partial{r^2}}+\frac{1}{r} \frac{\partial{}}{\partial{r}}-\frac{1}{r^2}-{\nabla_H^2}\right){u_r}\right]_{r=R},  
	\end{split}
	\end{equation} \\
	
	where $\boldsymbol{{\nabla_H}}=( \frac{\hat{\boldsymbol{\theta}}}{r}\frac{\partial}{\partial{\theta}}+\hat{\boldsymbol{z}} \frac{\partial}{\partial{z}})$ and $\nabla_H^2=\left(\frac{1}{r^2}\frac{\partial^2{}}{\partial{\theta^2}}+\frac{\partial^2{}}{\partial{z^2}}\right)$.
	
	\section{Solution of the linearized equations}
	Decomposing the velocity field $\boldsymbol u$ into irrotational component ${\boldsymbol \nabla}\phi$ and rotational component $\boldsymbol \chi$ as
	\begin{equation}\label{velocity}
	\boldsymbol u={\boldsymbol \nabla}\phi+\boldsymbol \chi.
	\end{equation}
	Linearized hydrodynamic equations (\ref{linear_navier_stokes}) and (\ref{continuity_2}) can be expressed as
	\begin{equation}\label{irrotational_navier_stokes}
	p+\rho \frac{\partial{\phi}}{\partial{t}}
	=0,
	\end{equation} 
	\begin{equation}\label{irrotational_continuity}
	\nabla^2 \phi =0,
	\end{equation}
	\begin{equation}\label{rotational_navier_stokes}
	\frac{\partial{\boldsymbol \chi}}{\partial{t}}+\nu [\boldsymbol \nabla \times \boldsymbol \nabla \times \boldsymbol{\chi}]=0,
	\end{equation}
	\begin{equation}\label{rotational_continuity}
	\boldsymbol \nabla \cdot \boldsymbol \chi=0,
	\end{equation}
	
	where kinematic viscosity ($\nu)=\mu/\rho$.\\
	
	Linearized pressure [Eq.~(\ref{irrotational_navier_stokes})] at the free surface can be written as
	\begin{equation}\label{pressure_surface_2}
	p|_{r=R}=-\rho \left(\frac{\partial{\phi}}{\partial{t}}\right)_{r=R}.
	\end{equation} 
	
	Equating Eq.~(\ref{pressure_surface_1}) and Eq.~(\ref{pressure_surface_2}) leads to 
	
	\begin{equation}\label{total_normal_stress_balace}
	\begin{multlined}
	\left(\frac{\partial{\phi}}{\partial{t}}\right)_{r=R}-G(t)\eta-{\frac{\sigma}{\rho}}\left(\frac{1}{R^2}
	+\frac{1}{R^2}\frac{\partial^2{}}{\partial{\theta^2}}
	+\frac{\partial^2{}}{\partial{z^2}}\right)\eta\\
	+2\nu{\frac{\partial{u_r}}{\partial{r}}}|_{r=R}=0.
	\end{multlined}
	\end{equation}
	
	\begin{figure*}[ht]
		\centering
		\subfigure[]{\includegraphics[width=0.48\textwidth]{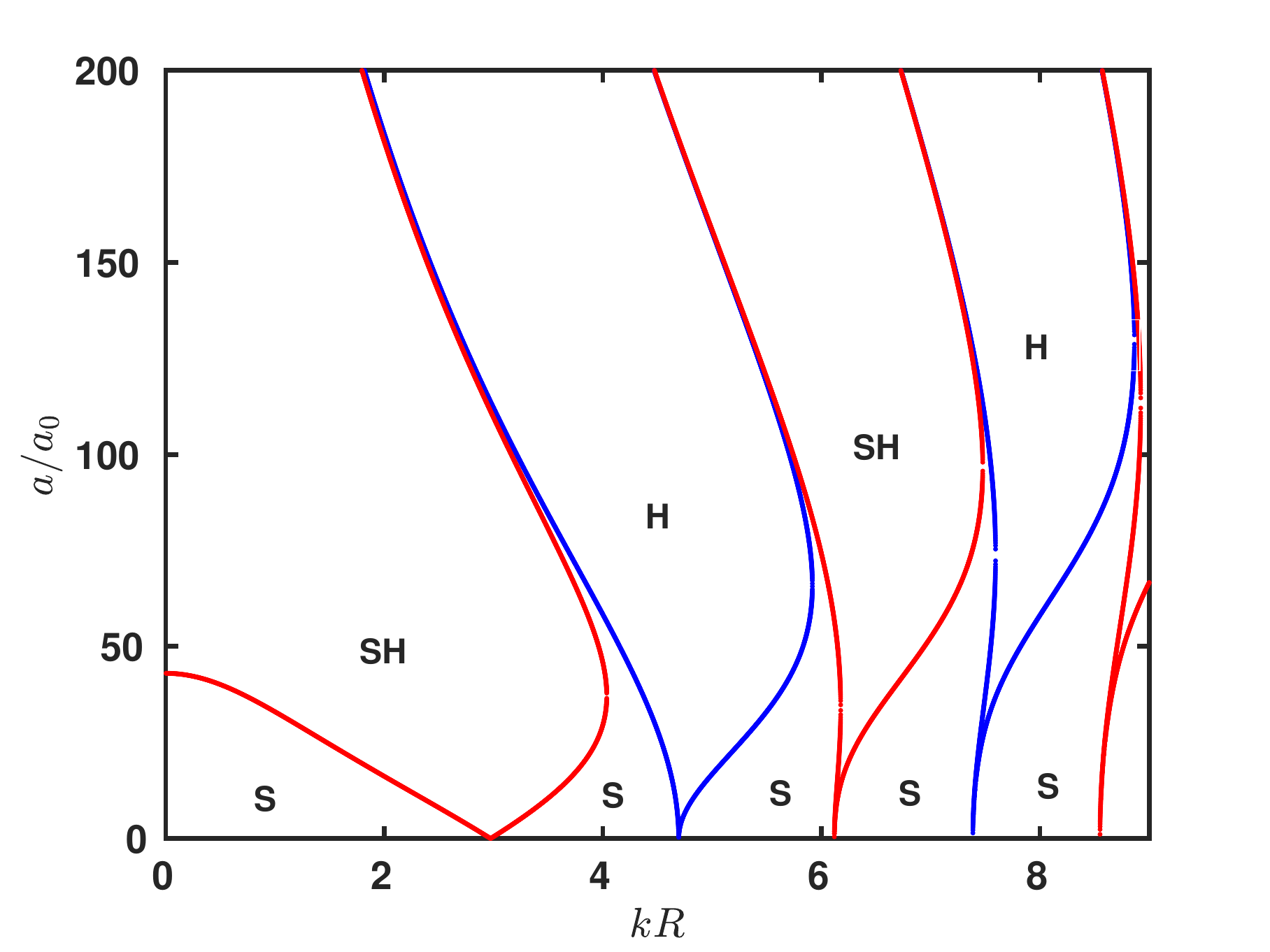}} 
		\subfigure[]{\includegraphics[width=0.48\textwidth]{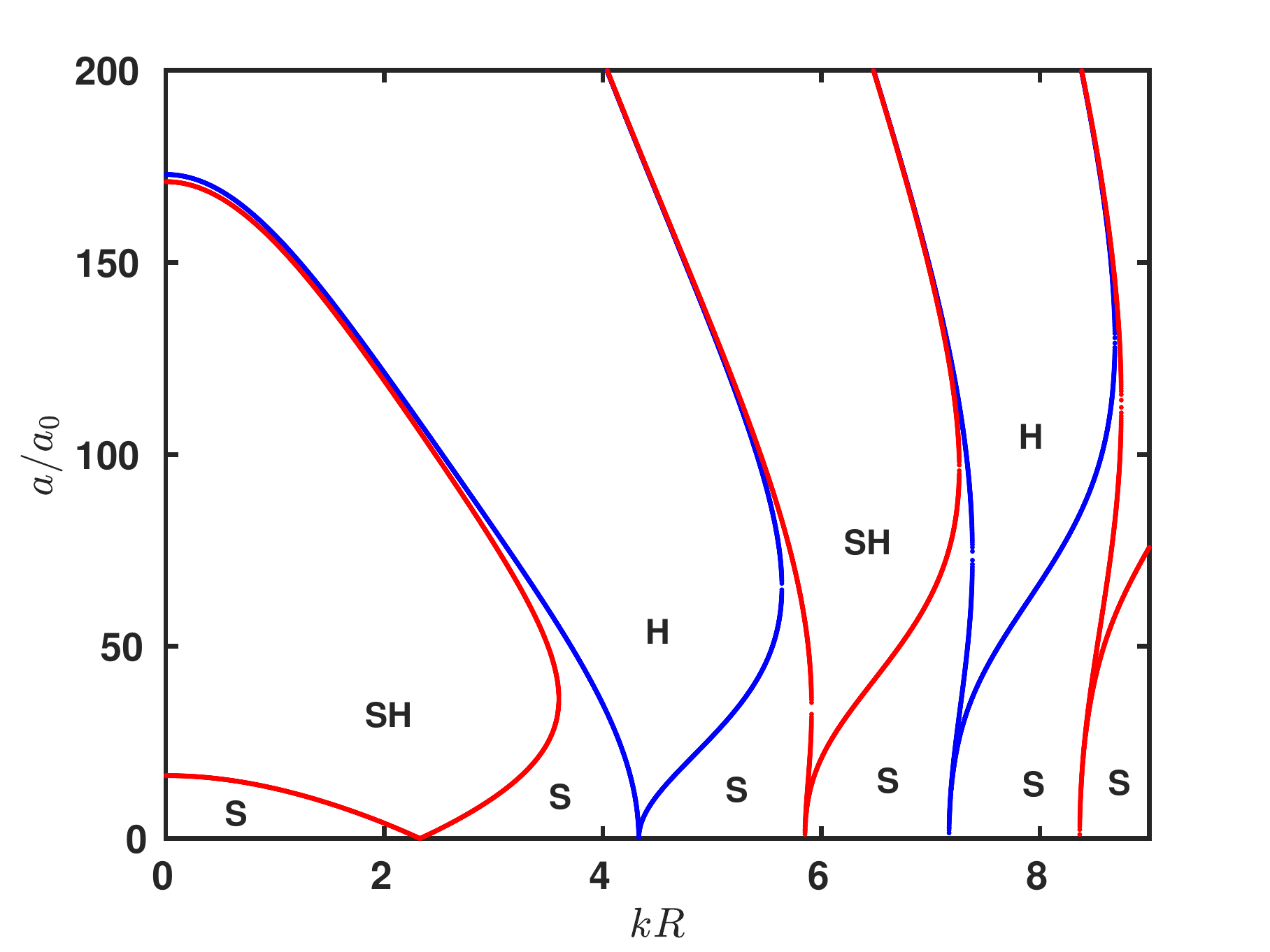}}\\ 
		\subfigure[]{\includegraphics[width=0.48\textwidth]{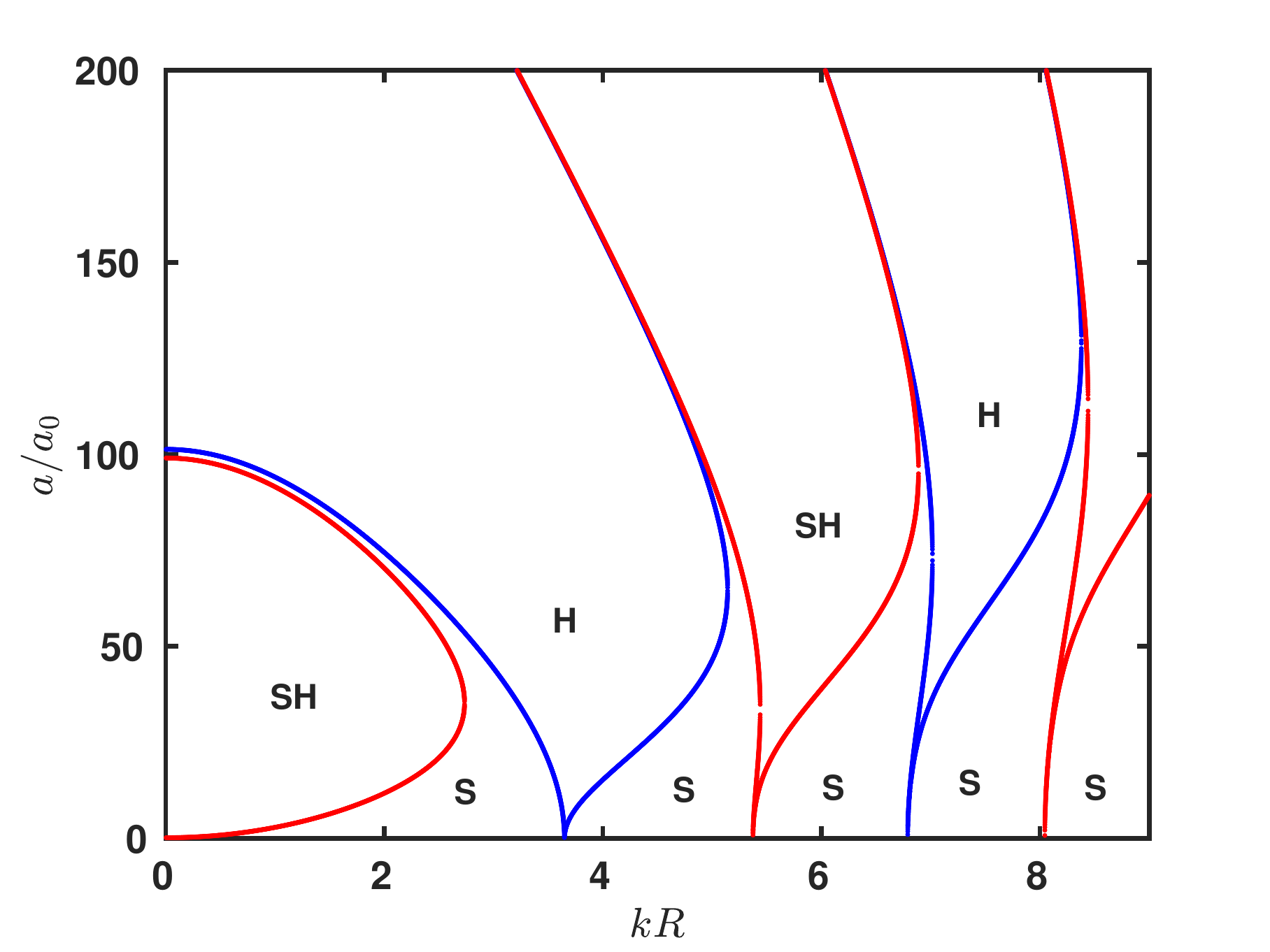}}
		\subfigure[]{\includegraphics[width=0.48\textwidth]{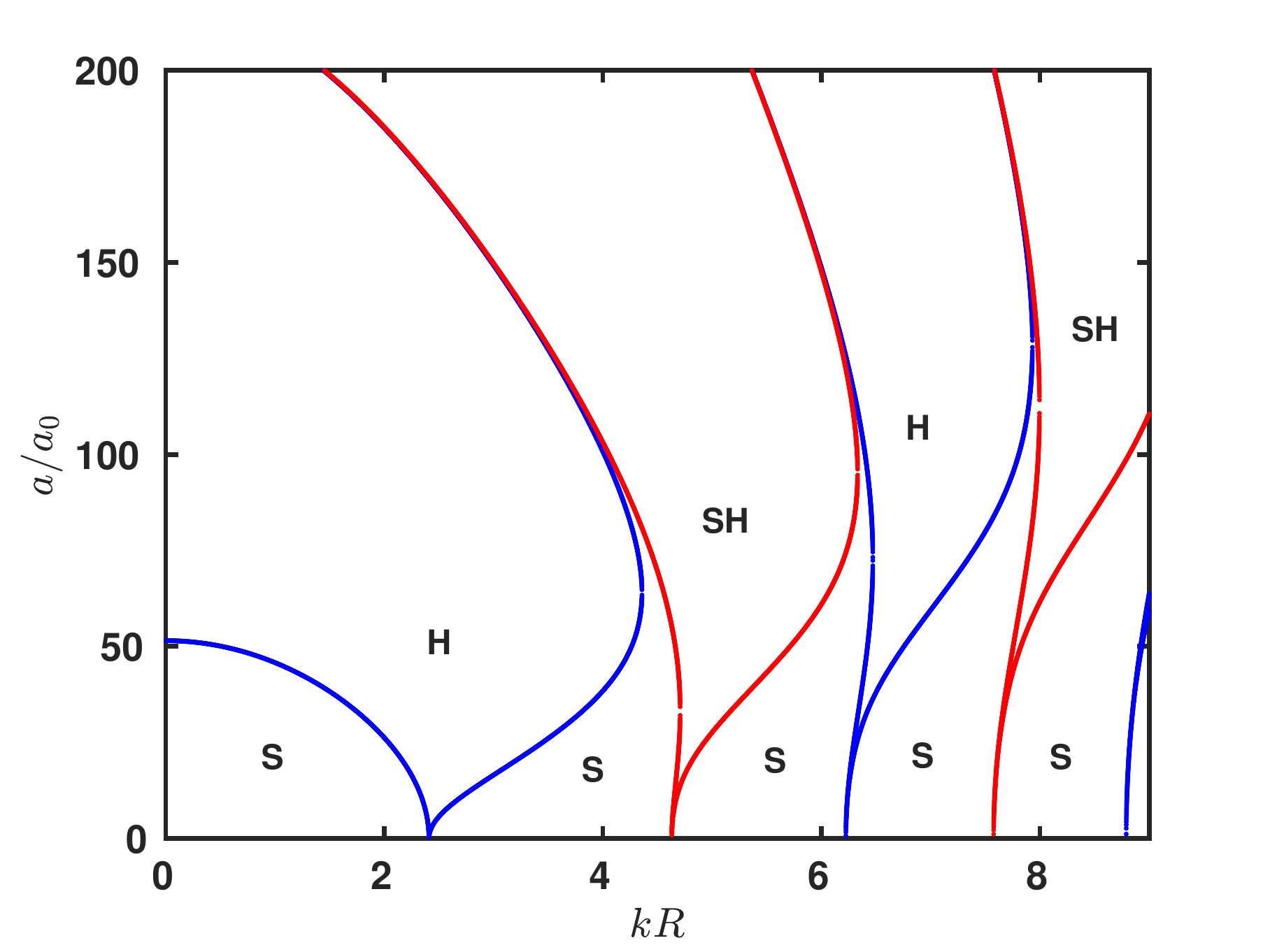}}
		\caption{(color online) Marginal stability boundaries of Mathieu equation [Eq.~(\ref{mathieu})]  are plotted for dimensionless forcing angular frequency, $\frac{\omega}{\omega_0}=9.73$. Red (gray) and blue (black) boundaries represent sub-harmonic (SH) and harmonic (H) case respectively. S represents stable region of the system. In the stability curves dimensionless forcing amplitude ($a/a_0$) are plotted with dimensionless axial wavenumber ($kR$) for (a) $m=1$ (b) $m=2$ (c) $m=3$ (d) $m=4$.}
		\label{fig2}
	\end{figure*}
	\begin{figure}[]
		\begin{center}
			\includegraphics[height=!, width=9 cm]{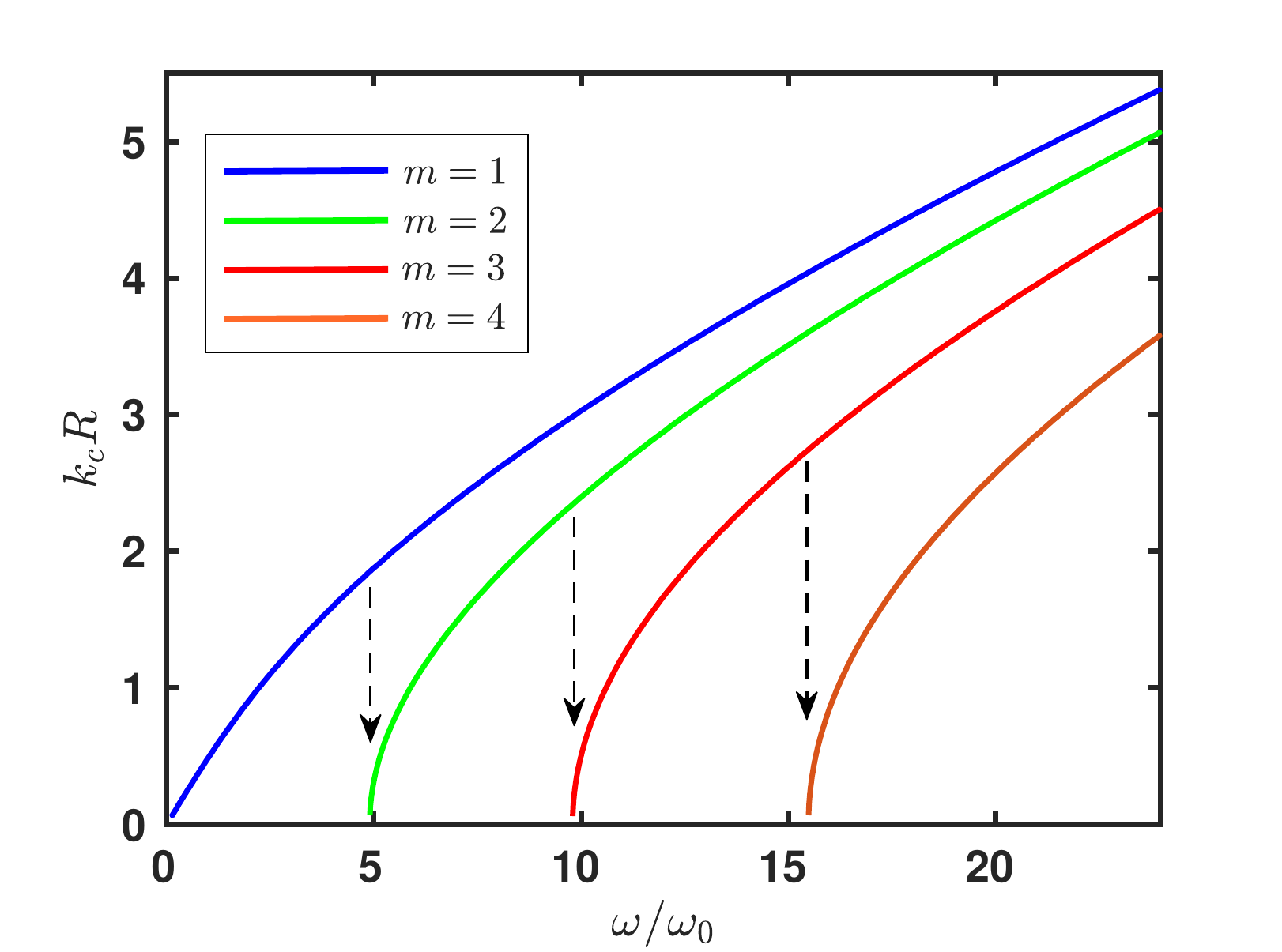}
			\caption{(Color online) {Dispersion curves for sub-harmonically excited standing waves on the cylindrical fluid surface for different azimuthal wavenumber (m). Blue, green, red, saffron colored curves correspond to $m = 1, 2, 3, 4$ respectively. Black arrow represents transition points of the patterns}
			} \label{fig3}
		\end{center}
	\end{figure}
	
	\subsection{Ideal fluid case}
	For the case of ideal fluid ($\nu=0$), Eq.~(\ref{total_normal_stress_balace}) reduces to 
	\begin{equation}\label{total_normal_stress_balace_ideal}
	\left(\frac{\partial{\phi}}{\partial{t}}\right)_{r=R}-G(t)\eta-{\frac{\sigma}{\rho}}\left(\frac{1}{R^2}
	+\frac{1}{R^2}\frac{\partial^2{}}{\partial{\theta^2}}
	+\frac{\partial^2{}}{\partial{z^2}}\right)\eta=0
	\end{equation}
	
	and continuity equation gives
	\begin{equation}\label{irrotational_continuity_ideal}
	\nabla^2 \phi =0.
	\end{equation}
	
	Considering periodic solution of Eq.~(\ref{irrotational_continuity_ideal}) in axial ($z$) and azimuthal direction ($\theta$), and using kinematic boundary condition, $\eta$ and $\phi$ can be expanded for a particular azimuthal wavenumber ($m$) as 
	\begin{equation}\label{expansions_eta_m_ideal}
	\eta_m=\bar{\eta}_m(t) e^{i(m\theta+kz)},
	\end{equation}
	\begin{equation}\label{expansions_phi_m_ideal}
	\phi_m=\frac{d\bar{\eta}_m(t)}{dt}  \frac{I_m(kr)}{kI'_m(kR)} e^{i(m\theta+kz)},
	\end{equation}
	where $I_m(kr)$  is the $m^{th}$ order modified Bessel function of first kind  and $I'_m(kr)$ is the first derivative of it. 
	Since at  $r=0$, $I_m$ is finite, $I_m$ is chosen in the expansion.

	For $m^{th}$ mode pattern, using the expansion of $\eta_m$ and $\phi_m$,  Eq.~(\ref{total_normal_stress_balace_ideal}) is converted to a Mathieu equation
	
	\begin{equation}\label{mathieu}
	\ddot{\bar\eta}_m+\omega_{m}^2 \left[ 1 + ( a/a_{m}) \cos{(\omega t)}\right]{\bar\eta}_m =0,
	\end{equation}
	
	where $\omega_{m}^2$ $=$ $\frac{\sigma}{\rho R^3}$$\left[\frac{kR I'_m(kR)}{I_m(kR)}(k^2R^2+m^2-1)\right]$ \\ 
	
	and $a_{m}$ $=$ $\frac{\sigma}{\rho R^2}$$\left[(k^2R^2+m^2-1)\right]$.\\

	To make dimensionless angular frequency and forcing amplitude, some $m$ independent constants are to be defined as follows

	\begin{equation}\label{omega_0}
	\omega_{0}^2=\frac{\sigma}{\rho R^3},
	\end{equation}
	\begin{equation}\label{a_0}
	a_{0}=\frac{\sigma}{\rho R^2}={\omega_{0}^2}R
	\end{equation}
	and
	\begin{equation}\label{a_0}
	\nu_{0}=\sqrt{\frac{\sigma R}{\rho}}={\omega_{0}}R^2.
	\end{equation}.

	As acceleration, $G(t)$ of the system is periodically varying with time; the stability problem can be analyzed using Floquet theory.     
	
	Floquet expansion~\cite{Kumar_1996} of $\bar{\eta}_m (t)$ can be expressed as
	\begin{equation}\label{Floquet:expansion}
	{\bar\eta}_m (t) = e^{(s+i\alpha \omega ) t} \sum_{n=-\infty}^{\infty} \eta_m^{(n)} e^{i n\omega t}.
	\end{equation}
	Here ($s+i\alpha\omega$) is Floquet exponent and $s$ is growth rate.\\
	The reality condition of surface deformation reads for harmonic case ($\alpha=0$) 
	\begin{equation}\label{reality_harmonics}
	\eta^{-n}_m =(\eta^n_m)^*
	\end{equation}
	and for sub-harmonic case ($\alpha=1/2$) \\
	\begin{equation}\label{reality_subharmonics}
	\eta^{-n}_m =(\eta^{n-1}_m)^*.
	\end{equation}
	
	Using Eq.~(\ref{Floquet:expansion}), Mathieu equation (\ref{mathieu}) leads to a difference equation
	\begin{equation}
	A_m^{(n)}\eta_m^{(n)} = a \left[ \eta_m^{(n-1)}+\eta_m^{(n+1)}\right],
	\end{equation}
	where 
	\begin{equation}
	A_m^{(n)} = -\frac{2 I_m(kR)}{kI'_m(kR)}\left[ \omega_m^2-(n+\alpha)^2 \omega^2\right].
	\end{equation}
	Time periodic acceleration couples different temporal modes in the system. This recursion relation can be converted to a eigenvalue matrix equation as
	\begin{equation}
	A_m\eta_m = aB \eta_m,
	\end{equation}
	where $A_m$ is a diagonal square matrix and B is a banded square matrix with two sub-diagonals.
	
	To solve this eigenvalue problem, truncation is needed somewhere on the temporal modes. Reality conditions suggest that for the subharmonic case, the truncation should be made by the even number of temporal modes, and for the harmonic case, it should be made by the odd number of temporal modes. Real eigenvalues of the matrix can be determined as a function of the axial wavenumber $k$ for fixed values of $\omega$ and $m$. These give marginal (the growth rate, $s =0$) stability boundaries (which looks like a tongue in the parameter space of $a$ and $k$) without interpolation.
	
	The marginal stability boundaries have shown in Fig. \ref{fig2} for four different $m$ at dimensionless forcing frequency $\omega/\omega_{0}=9.73$ in the parameter space of dimensionless axial wavenumber ($kR$) and forcing amplitude ($a/a_0$). All the figures in  Fig. \ref{fig2}(a-d) show family of sub-harmonic (red color) and harmonic (blue color) tongues for $m=1,2,3,4$. To solve this problem, fifty temporal modes are taken for the subharmonic case and fifty-one temporal modes for the harmonic case. Inside every tongue which originates from zero threshold acceleration ($a/a_0=0$) at critical wavenumber ($k_cR$), the cylinder becomes unstable by the spatial shape~\cite{Dasgupta_2} corresponding to $m$  and oscillates at angular frequencies
	\begin{equation}
	\omega_m=\frac{n}{2} \omega,
	\end{equation}
	where $n=1,2,3 ...$ represents first subharmonic, first harmonic and second sub-harmonic tongue etc., respectively. Outside region of the tongue the system remains stable. In every tongue [Fig.~\ref{fig2}(a)-(b)] of the patterns which correspond to $m=1$ and $2$, there are two values of $kR$ for a single value of $a/a_0$; but in Fig.~\ref{fig2}(c) which corresponds to $m=3$, at first subharmonic tongue $a/a_0$ is single valued function of $kR$. For $m=4$ first subharmonic tongue [Fig.~\ref{fig2}(d)] disappears. This indicates that at this forcing frequency ($\omega/\omega_0=9.73$) only subharmonic pattern with $m=1$ and $2$ can be exited on the cylinder.   
	
	In Fig. \ref{fig3}, the variations of the axial critical wavenumber, i.e., the wavenumber correspond to a minimum ($a=0$) of the first sub-harmonic tongue, are shown with forcing frequency for four different $m$. From the dispersion curves, shown in Fig. \ref{fig3}, it is clear that at $\omega/\omega_{0}=9.73$ sub-harmonically excited standing waves can be excited only for $m=1, 2$ and the same can not be exited for $m>2$. Vertical arrows in the dispersion curves indicate the transition point of patterns from $m$ to $m+1$. These types of transitions are observed in our recent experiment~\cite{Dilip}.
	
	\begin{figure*}[ht]
		\centering
		\subfigure[]{\includegraphics[width=0.48\textwidth]{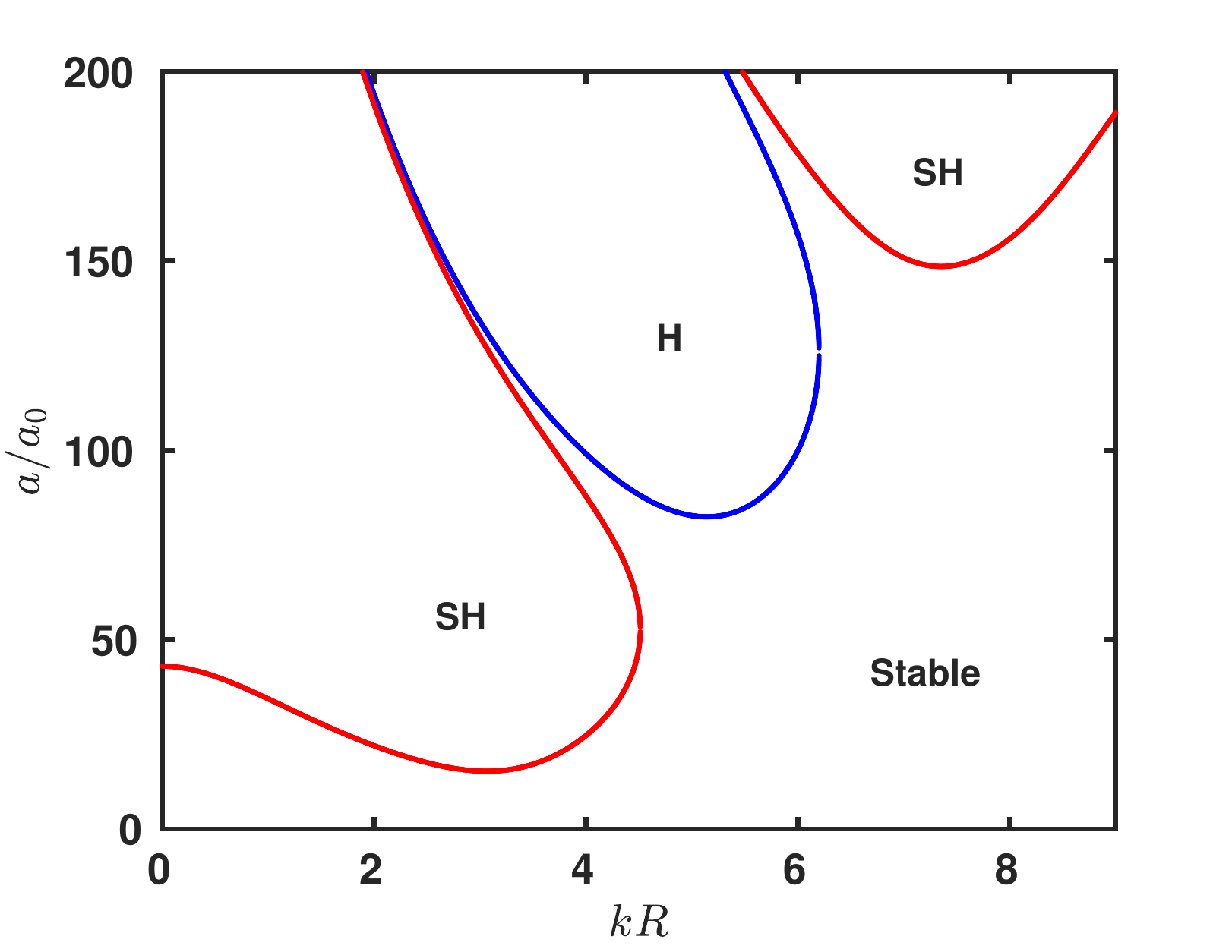}} 
		\subfigure[]{\includegraphics[width=0.48\textwidth]{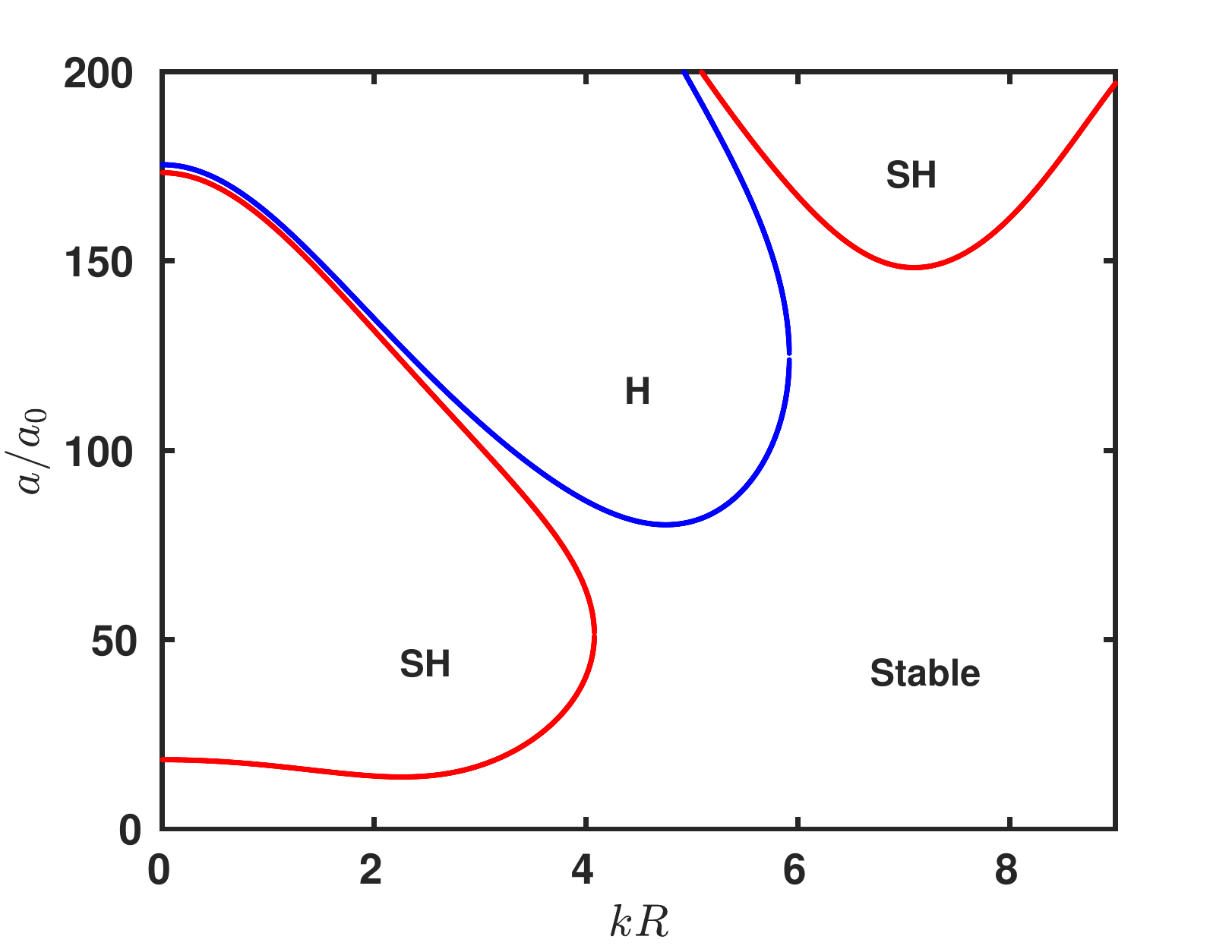}}\\
		\subfigure[]{\includegraphics[width=0.48\textwidth]{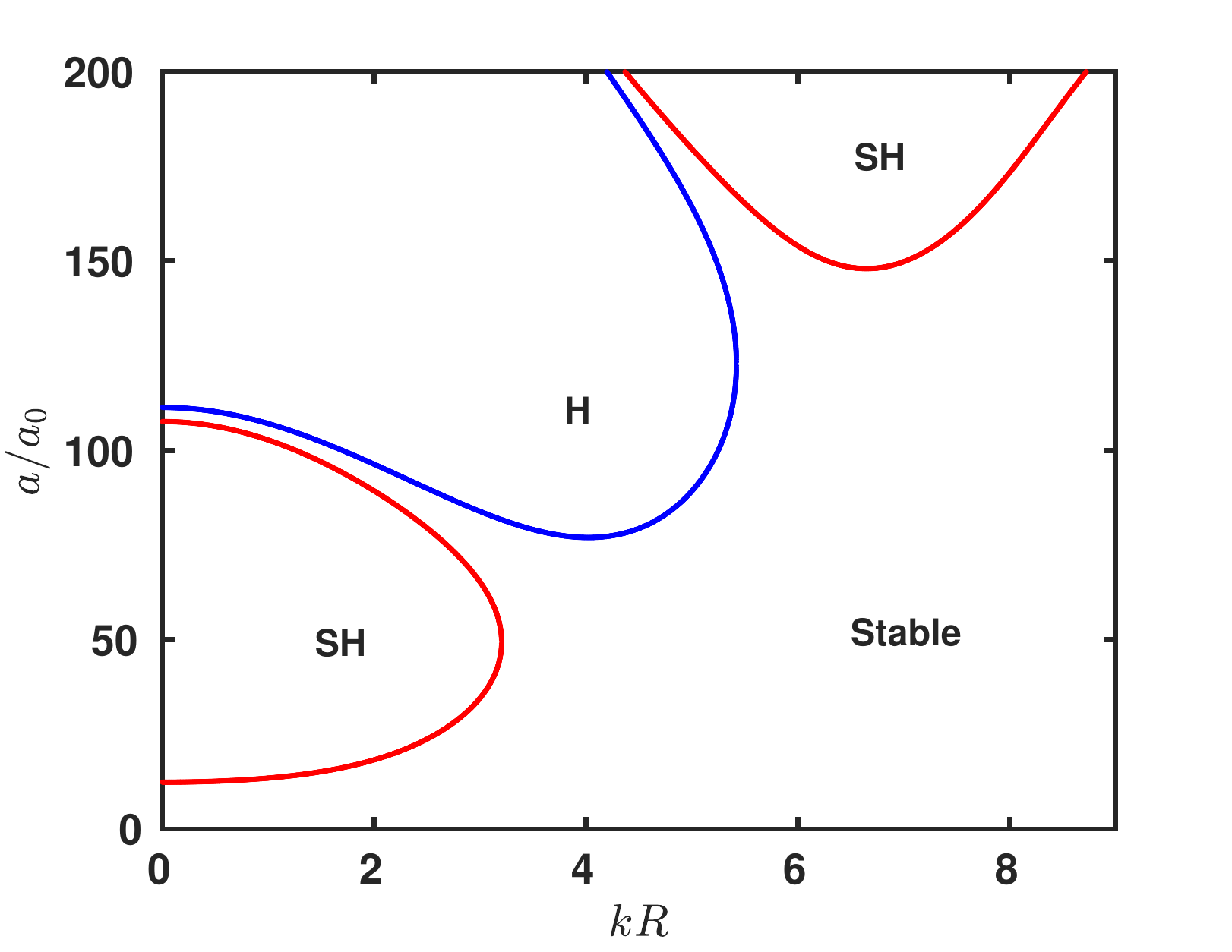}}
		\subfigure[]{\includegraphics[width=0.48\textwidth]{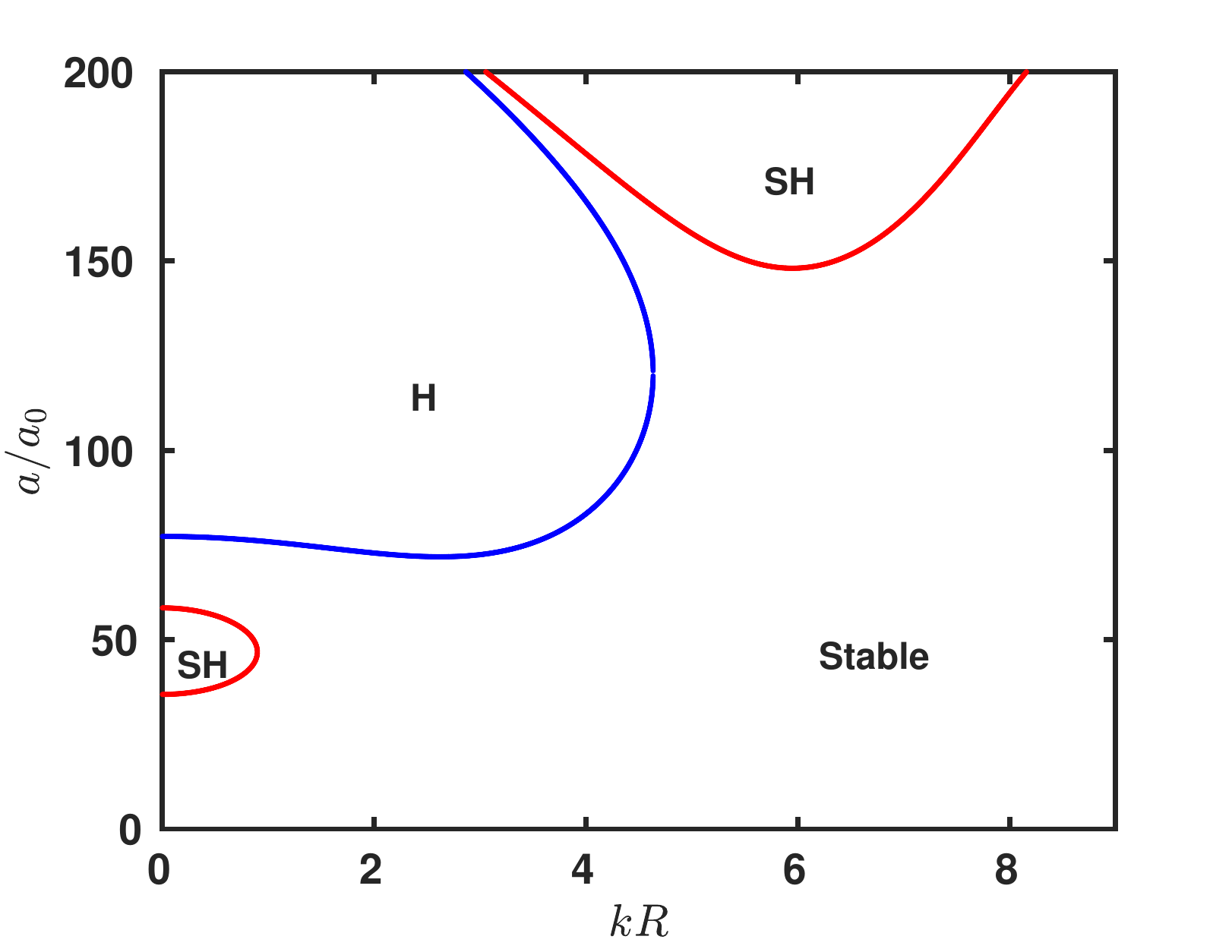}}
		\caption{(color online) Instability tongues of damped Mathieu equation [Eq.~(\ref{damped_mathieu_equation})] are plotted for (a) $m=1$ (b) $m=2$ (c) $m=3$ (d) $m=4$. Radial acceleration of the cylinder which is formed with a fluid whose dimensionless viscosity is $\nu/\nu_0=0.21$, is modulated time periodically at $\frac{\omega}{\omega_0}=9.73$.  Red (gray) and blue (black) boundaries are represents the sub-harmonic (SH) and the harmonic (H) case respectively. Stable represents stable state of the system. }
		\label{fig4}
	\end{figure*}
	
	\subsection{Viscous fluid case}
	Now, consider the rotational component of the velocity as 
	\begin{equation}\label{chi_1}
	\boldsymbol \chi =\boldsymbol \nabla \times \boldsymbol \nabla \times (\psi\hat{\boldsymbol{r}}).
	\end{equation}
	
	So, the velocity potentials $\phi$ and $\psi$, and the surface deformation $\eta$  for the pattern with azimuthal wavenumber $m$ can be expanded as:
	\begin{equation}\label{expansions_eta_m}
	\eta_m(\theta,z,t)=\sum_{n=-\infty}^{\infty}\eta_m^{(n)} e^{i(m\theta+kz)+\zeta_n t},
	\end{equation}
	\begin{equation}\label{expansions_phi_m}
	\phi_m(r,\theta,z,t)=\sum_{n=-\infty}^{\infty} B_m^{(n)} I_m(kr) e^{i(m\theta+kz)+\zeta_n t},
	\end{equation}
	\begin{equation}\label{expansions_psi_m}
	\psi_m(r,\theta,z,t)=\sum_{n=-\infty}^{\infty}D_m^{(n)} r I_m(lr)e^{i(m\theta+kz)+\zeta_n t}.
	\end{equation}
	
	Floquet exponent, $\zeta_n=s+i(n+\alpha)\omega$ and $l^2=k^2+\zeta_n/\nu$. $s$ corresponds to growth rate,  $\alpha=0$ for harmonic response and $\alpha=1/2$ for subharmonic response.
	
	Radial velocity component can be written with the help of the velocity potentials, $\phi, \psi$ as
	
	\begin{equation}\label{radial_velocity}
	\begin{split}
	u_r & =\frac{\partial \phi}{\partial r}-\left(\frac{1}{r^2}\frac{\partial^2{}}{\partial{\theta^2}}
	+\frac{\partial^2{}}{\partial{z^2}}\right)\psi\\
	& =\sum_{n=-\infty}^{\infty}L_m^{(n)}(r) e^{i(m\theta+kz)+\zeta_n t}, \\ 
	\end{split}
	\end{equation} \\
	where $L_m^{(n)}=[B_m^{(n)} k {I'_m}(kr)+rD_m^{(n)} (\frac{m^2}{r^2}+k^2) I_m(lr)].$\\
	
	Using Eq.~(\ref{linearised_kinematic}) and Eq.~(\ref{horizontal_divergence_tengential_stress_balance}), I get
	
	\begin{equation}\label{B_m}
	B_m^{(n)}=\frac{-\eta_{m}^{(n)}\zeta_n R \delta_{1m}}{k\delta_{2m}},
	\end{equation}
	
	\begin{equation}\label{D_m}
	D_m^{(n)}=\frac{\eta_{m}^{(n)}\zeta_n \delta_{3m}}{R\delta_{2m}},
	\end{equation}

	$\delta_{1m}=[R(m^2/R^2+k^2)^2{I_m}(lR)+4k^2l{I'_m}(lR)-l(m^2/R^2+k^2){I'_m}(lR)+l^2R(m^2/R^2+k^2){I''_m}(lR)]$,\\
	
	$\delta_{2m}=[-(m^2/R^2+k^2){I'_m}(kR){I_m}(lR)
	+kR(m^2/R^2+k^2){I''_m}(kR){I_m}(lR)+k^2R^2(m^2/R^2+k^2){I'''_m}(kR){I_m}(lR)-4k^2lR{I'_m}(kR){I'_m}(lR)+lR(m^2/R^2+k^2){I'_m}(kR){I'_m}(lR)-l^2R^2(m^2/R^2+k^2){I'_m}(kR){I''_m}(lR)]$,\\
	
	$\delta_{3m}=[-{I'_m}(kR)+kR{I''_m}(kR)+k^2R^2{I'''_m}(kR)\\
	+R^2(m^2/R^2+k^2){I'_m}(kR)]$.\\
	
	Here,  $I{''}_m$ and $I{'''}_m$ are the second and third-order derivative of $m^{th}$ order modified Bessel function of the first kind.
	
	Using Eq.~(\ref{expansions_eta_m}), (\ref{expansions_phi_m}) and  (\ref{radial_velocity}) for the $m^{th}$ mode pattern, Eq.~(\ref{total_normal_stress_balace}) can be expressed as
	\begin{equation}\label{total_normal_stress_balace_2}
	\begin{multlined}
	\sum_{n=-\infty}^{\infty}[\eta_{m}^{(n)}\zeta_n^2 B_{0m}^{(n)} I_m(kR)
	-G(t)\eta_{m}^{(n)}\\
	+{\frac{\sigma}{\rho}}\left(\frac{m^2}{R^2}-\frac{1}{R^2}+k^2\right)\eta_{m}^{(n)}\\
	+\eta_{m}^{(n)}\zeta_n C_{0m}^{(n)} \nu] e^{i(m\theta+kz)+\zeta_n t}=0.
	\end{multlined}
	\end{equation}
	
	Where
	$B_{0m}^{(n)}=\frac{B_m^{(n)}}{\eta_{m}^{(n)} \zeta_n}$,
	$D_{0m}^{(n)}=\frac{D_m^{(n)}}{\eta_{m}^{(n)} \zeta_n}$,\\

	$C_{0m}^{n}=2[B_{0m}^{(n)} k^2 I''_m(kR)+D_{0m}^{(n)}(-m^2/R^2+k^2)I_m(lR)+D_{0m}^{(n)}lR(m^2/R^2+k^2)I'_m(lR)]$.\\

	Eq.~(\ref{total_normal_stress_balace_2}) can be written as 
	\begin{equation}\label{damped_mathieu_equation}
	\begin{multlined}
	B_{0m}^{(n)} I_m(kR) \frac{d^2\eta_m}{dt^2}+\nu C_{0m}^{(n)}  \frac{d\eta_m}{dt} \\
	+\left[-G(t)+{\frac{\sigma}{\rho}}\left(\frac{m^2}{R^2}-\frac{1}{R^2}+k^2\right)\right]\eta_{m}
	=0.
	\end{multlined}
	\end{equation}
	
	Equation~(\ref{damped_mathieu_equation}) is a standard damped Mathieu equation.
	
	Equation~(\ref{total_normal_stress_balace_2}) leads to a difference equation ~\cite{Kumar_1996}
	\begin{equation}\label{total_normal_stress_balace_6}
	\begin{multlined}
	A_m^{(n)}\eta_m^{(n)}=a[\eta_m^{(n-1)}+\eta_m^{(n+1)}],
	\end{multlined}
	\end{equation}
	
	where
	$A_m^{(n)}=\\
	-2\left[\zeta_n^2B_{0m}^{(n)}I_m(kR)+{\frac{\sigma}{\rho}}({m^2}/{R^2}-{1}/{R^2}+k^2)+\zeta_n C_0 \nu \right].$\\
	
	This eigenvalue problem is also solved in the same way, which is discussed in the ideal fluid case. To solve the eigenvalue problem, fifty temporal modes are taken for the subharmonic case and fifty-one temporal modes for the harmonic case. This eigenvalue problem is solved in MATLAB. The code is validated by the results of Kumar's~\cite{Kumar_1996} work.  In Fig.~\ref{fig4}(a-d), dimensionless forcing amplitudes of the viscous cylinder ($\nu/\nu_0=0.21$) are plotted with dimensionless axial wavenumber for $m=1,2,3,4$, respectively at $\omega/\omega_0=9.73$. Due to consideration of viscous dissipation, extra term added to the Mathieu equation.
	\begin{figure}[h]
		\centering
		\subfigure[]{\includegraphics[width=0.48\textwidth]{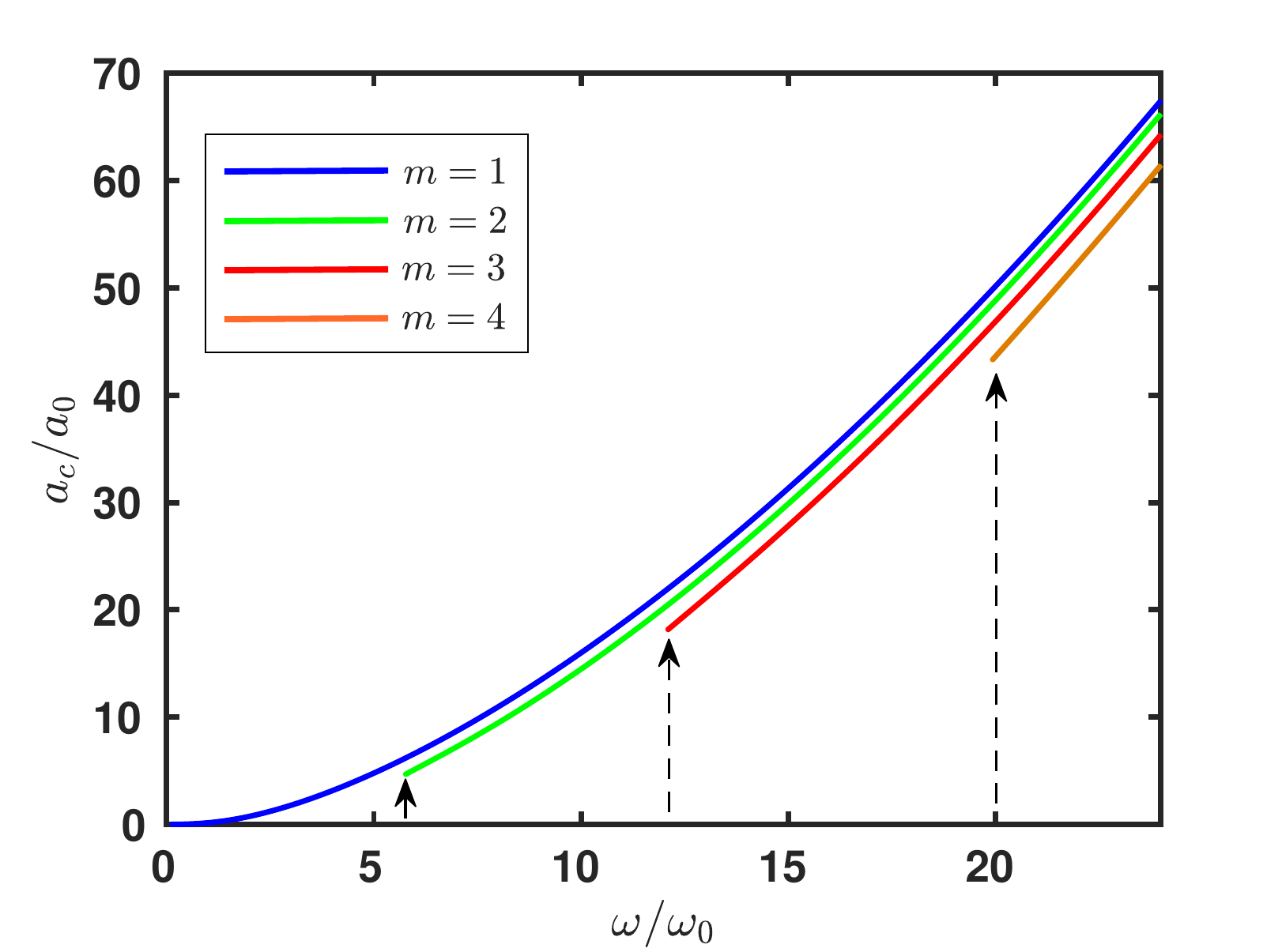}} 
		\subfigure[]{\includegraphics[width=0.48\textwidth]{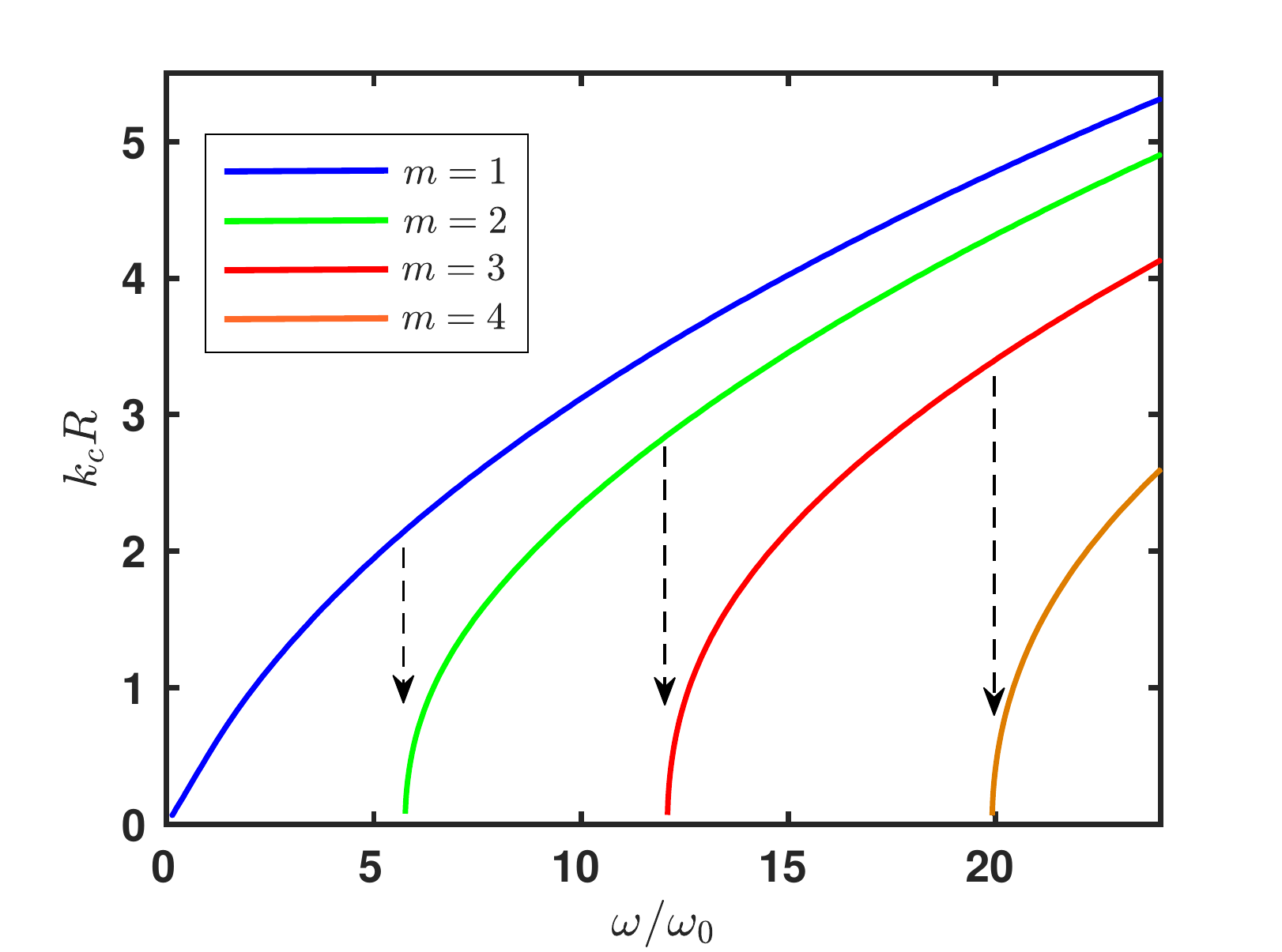}}\\ 
		\caption{(color online) Variations of critical forcing amplitude with forcing frequency [(a)] and dispersion curves [(b)] for four different sub-harmonically excited standing wave patterns ($m$) on the viscous cylinder. Dimensionless viscosity of the fluid is considered $\nu/\nu_0=0.21$.}
		\label{fig5}
	\end{figure}
	Unlike the ideal fluid case, instability tongues do not touch on $a/a_0=0$ axis; another way to say, the critical value of forcing amplitude is not zero due to the major role of viscosity. Due to the effect of viscosity, shape, size, and position of the tongues, changes significantly. This indicates that to overcome viscous force, unlike the inviscid case, a certain amount of extra energy is required to unstable the system even at critical axial wavenumber. Another important consequence of the viscous effect is that the critical axial wavenumber also changes significantly, i.e., the critical point of the instability tongue shifts in both vertical and horizontal direction. Also, for the same reason, which is discussed in the ideal fluid case,  at $\omega/\omega_0=9.73$, only the pattern with $m=1,2$ can be excited on the surface.
	
	\begin{figure}[h]
		\centering
		\subfigure[]{\includegraphics[width=0.48\textwidth]{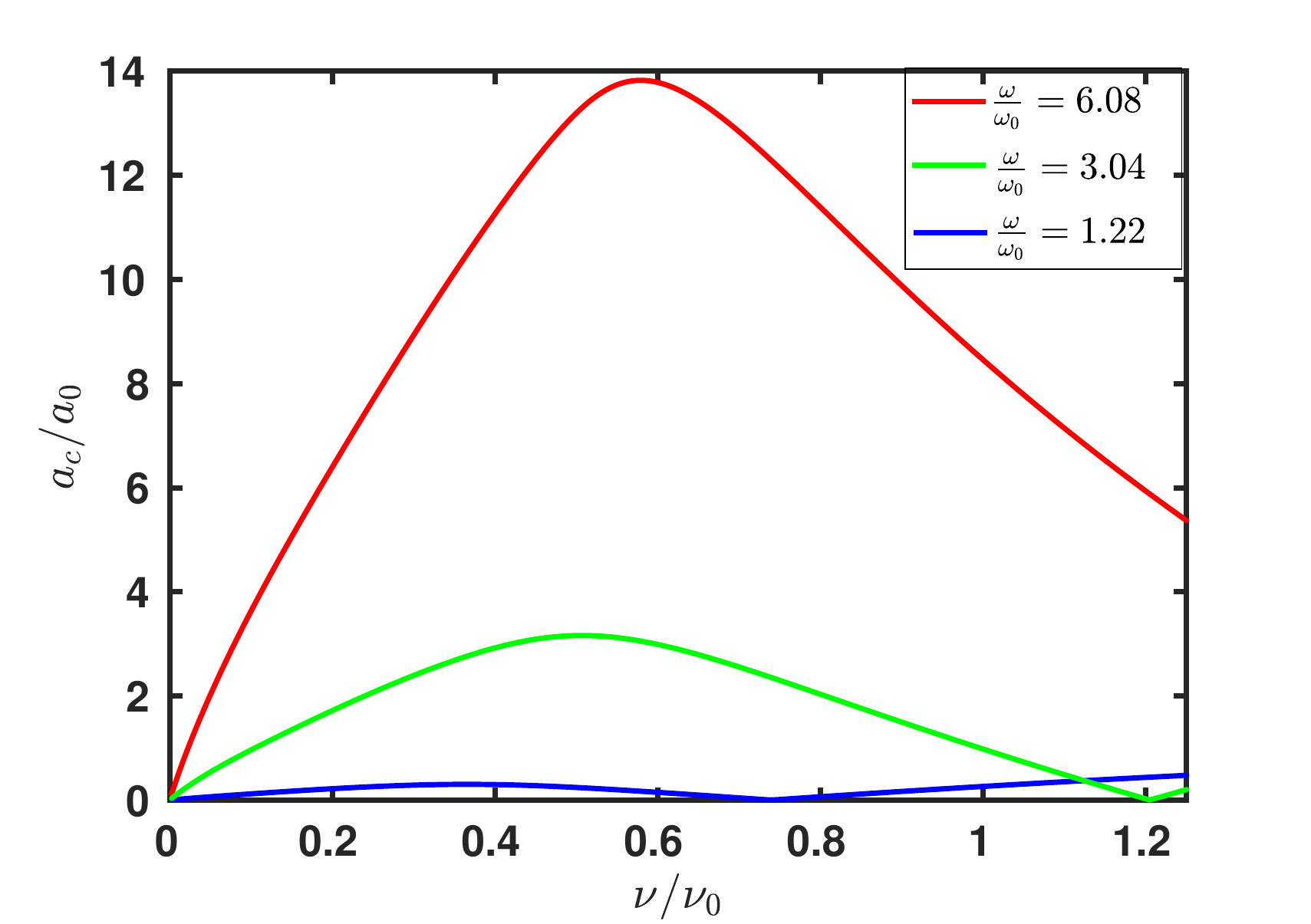}} 
		\subfigure[]{\includegraphics[width=0.48\textwidth]{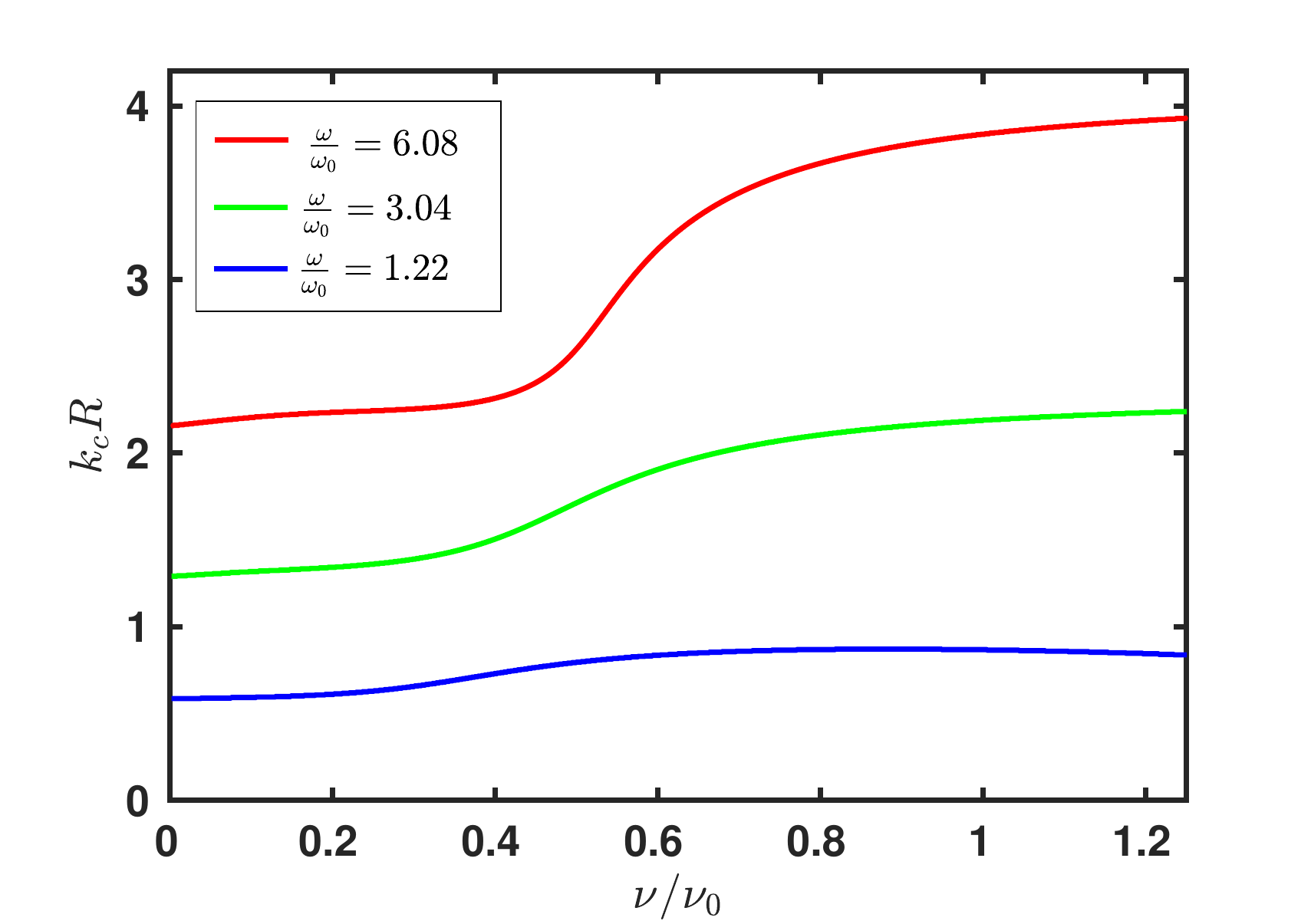}}\\ 
		\caption{(color online) Viscosity dependence of dimensionless threshold amplitude $(a_c/a_0)$ [(a)] and dimensionless critical axial wavenumber $(k_cR)$ [(b)] for sub-harmonically excited standing wave patterns with $m=1$ at three different forcing frequency.}
		\label{fig6}
	\end{figure}
	In Fig.~\ref{fig5}(a), variations of the critical forcing amplitude with forcing frequency are shown for $\nu/\nu_0=0.21$ to appreciate significant effect of the viscosity. Like ideal fluid case, the arrows are showing the transition points of the patterns. In Fig. \ref{fig5}(b), $k_cR$ is plotted with $\omega/\omega_0$ to show the viscous effects on the dispersion relation. Other parameters in Fig.~\ref{fig5}(b) are same as Fig.~\ref{fig5}(a). By comparing dispersion curves of the ideal case (Fig.~\ref{fig3}) with the viscous case [Fig.~\ref{fig5}(b)], it is clear that viscosity do have a major impact at higher azimuthal wavenumber ($m$), which is observed for viscous spherical droplet~\cite{Adou_Tuckerman_2016} also.
	
	Fig.~\ref{fig6}(a-b) show variations of the critical amplitude and the critical axial wavenumbers, respectively, with the viscosity at different forcing frequency for $m=1$. The variation of the critical amplitude is strongly non-monotonic. It increases to a maximum value, then comes to nearly zero, and finally, it again increases. This type of non-monotonic behavior of the critical amplitude is new in the Faraday instability. After initial flatness, $k_cR$ starts increasing sharply with viscosity and then nearly saturates at higher viscosity. The reason behind the increase of $k_cR$ at a lower viscosity is explained by Kumar and Tuckerman~\cite{Kumar_1994}.
	\begin{figure}[h!]
		\centering
		\subfigure[]{\includegraphics[width=0.48\textwidth]{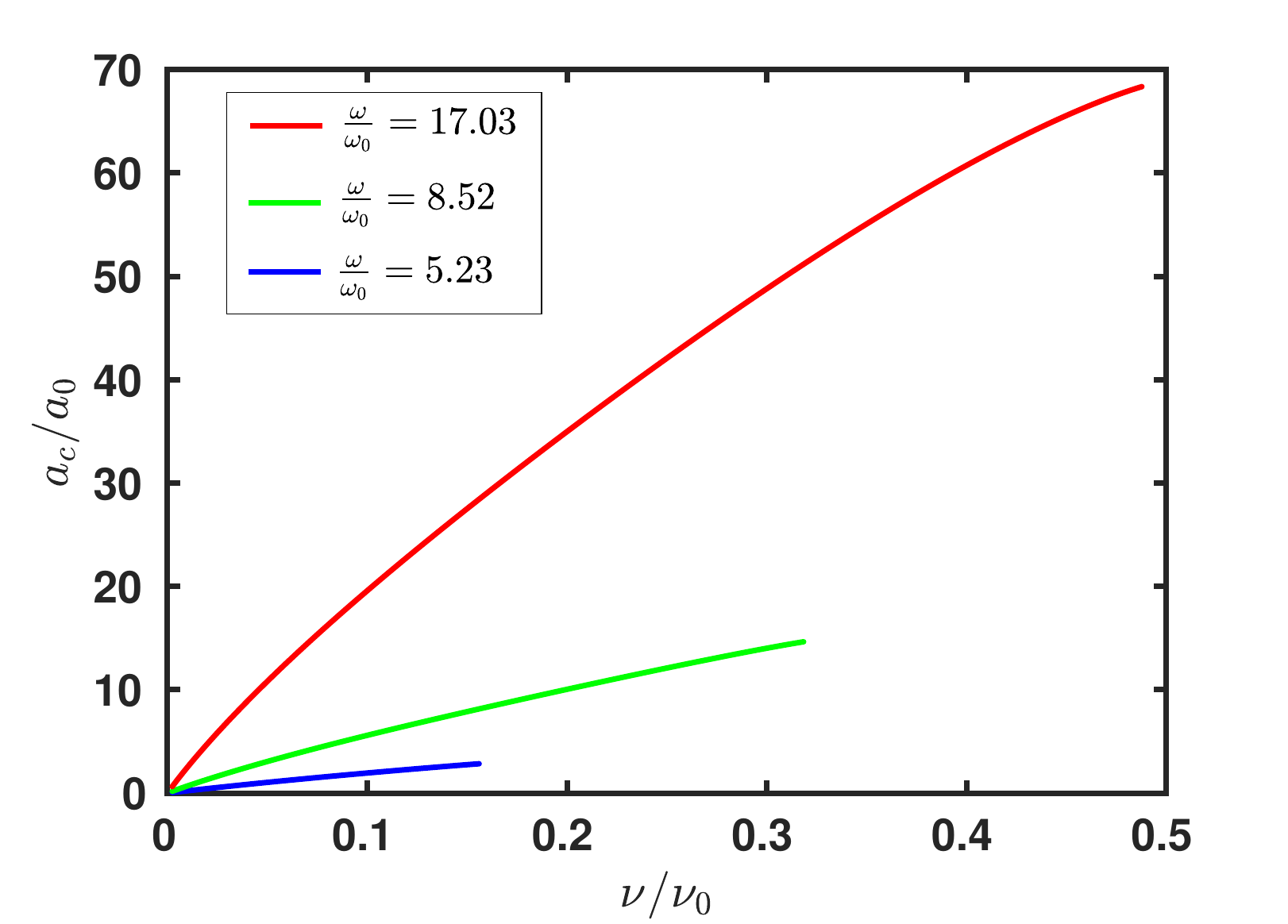}} 
		\subfigure[]{\includegraphics[width=0.48\textwidth]{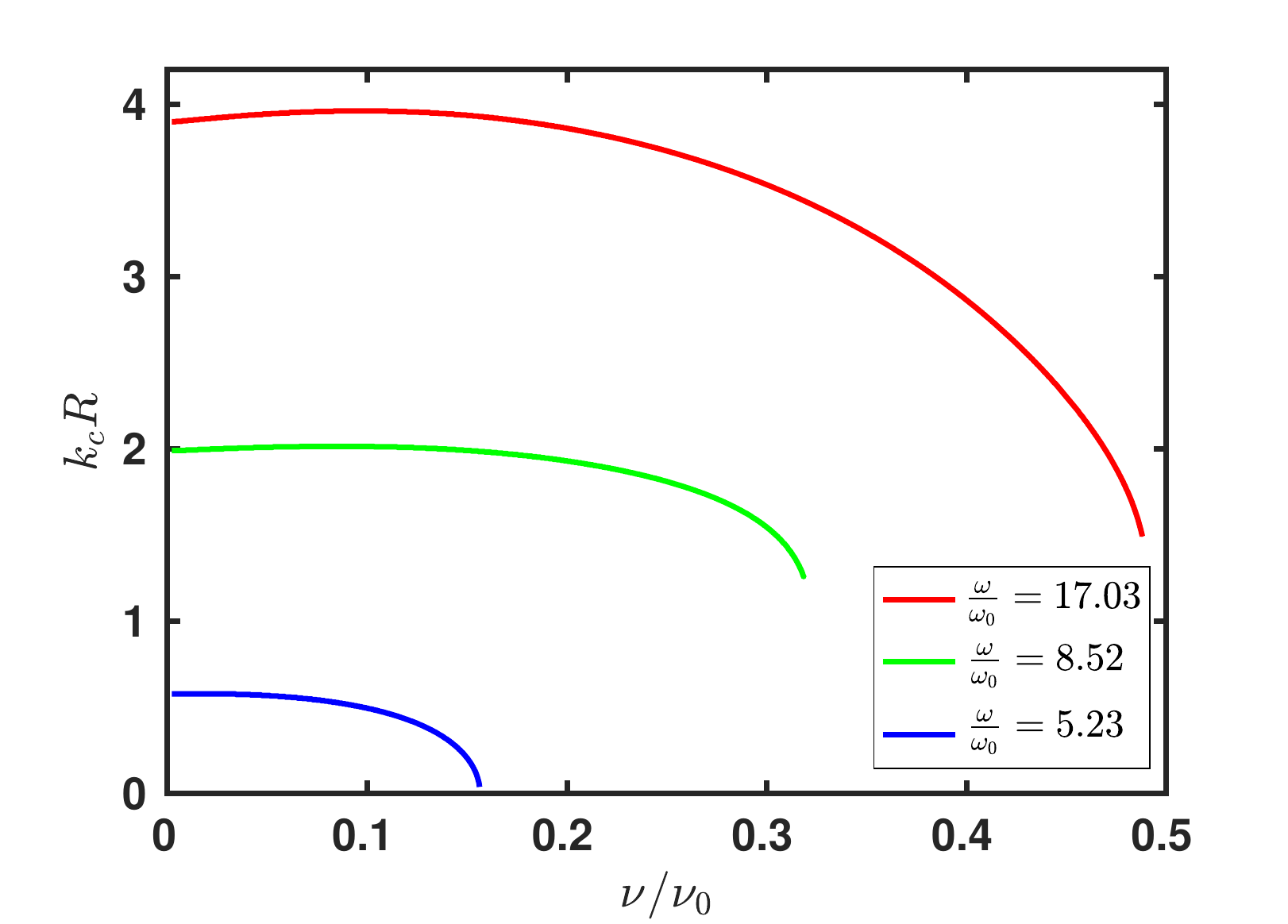}}\\ 
		\caption{(color online) Viscosity dependence of $(a_c/a_0)$ [(a)] and $(k_cR)$ [(b)] for sub-harmonically excited standing waves pattern with $m=2$ at three different forcing frequency.}
		\label{fig7}
	\end{figure}
	
	\begin{figure*}[t]
		\centering
		\subfigure[]{\includegraphics[width=0.48\textwidth]{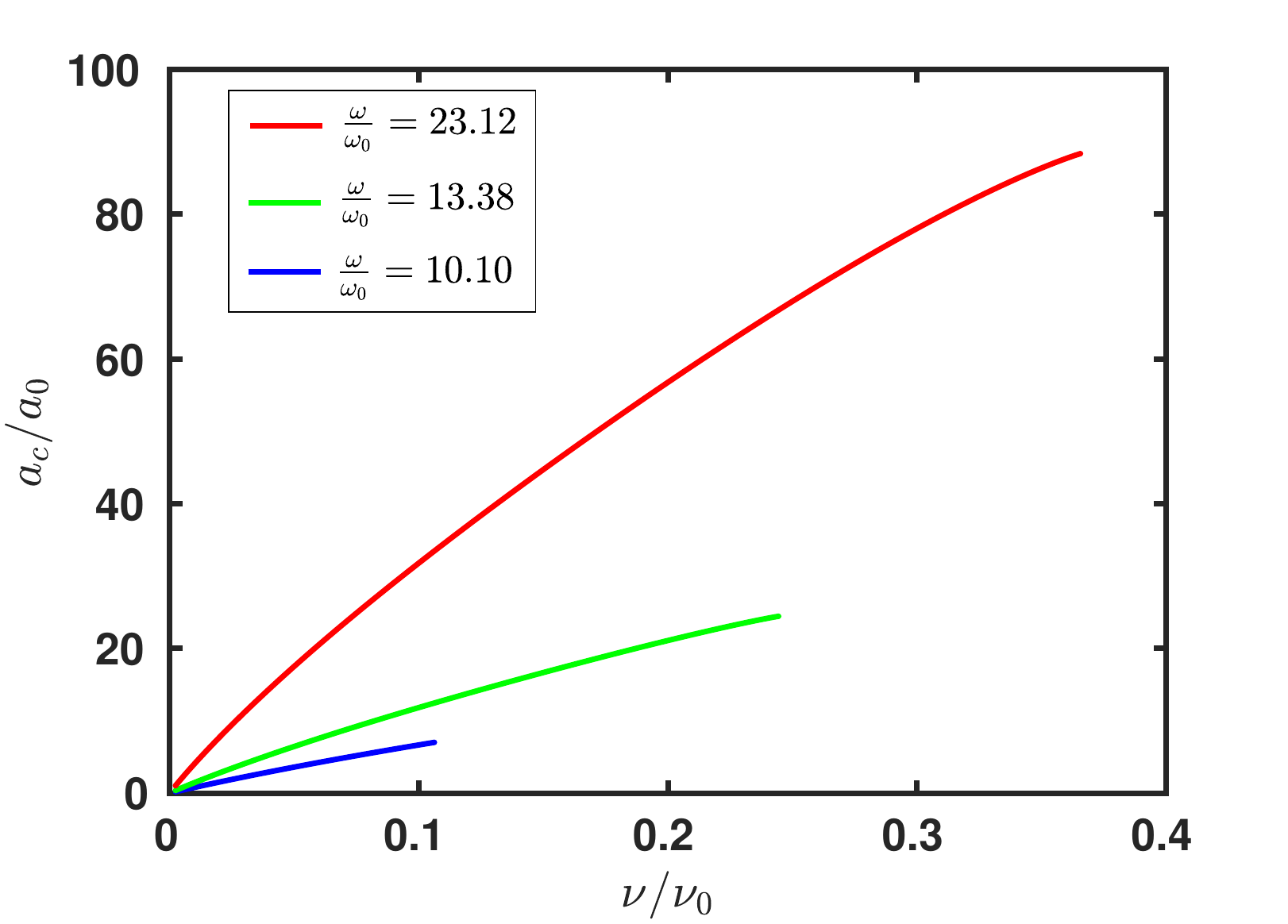}} 
		\subfigure[]{\includegraphics[width=0.48\textwidth]{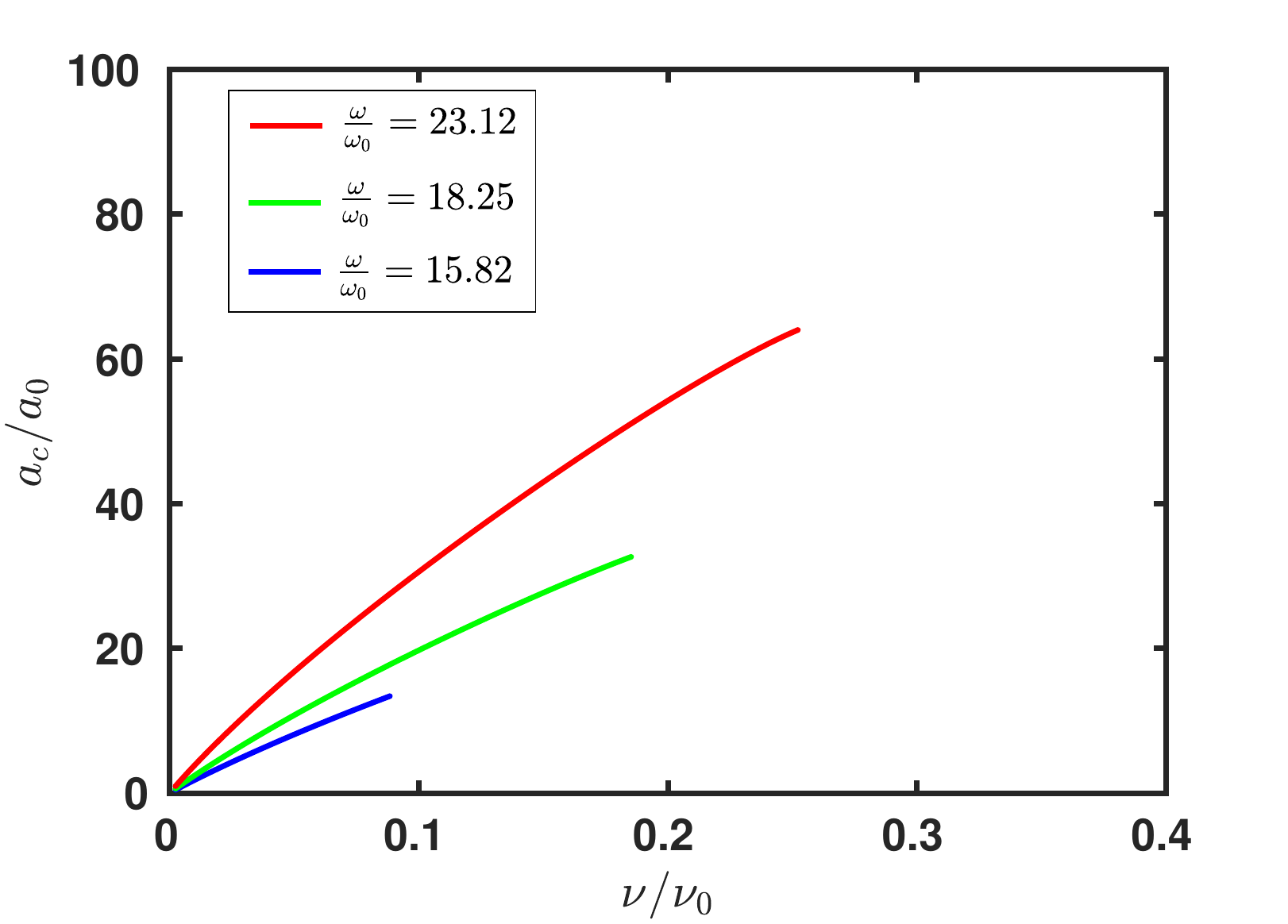}}
		\subfigure[]{\includegraphics[width=0.48\textwidth]{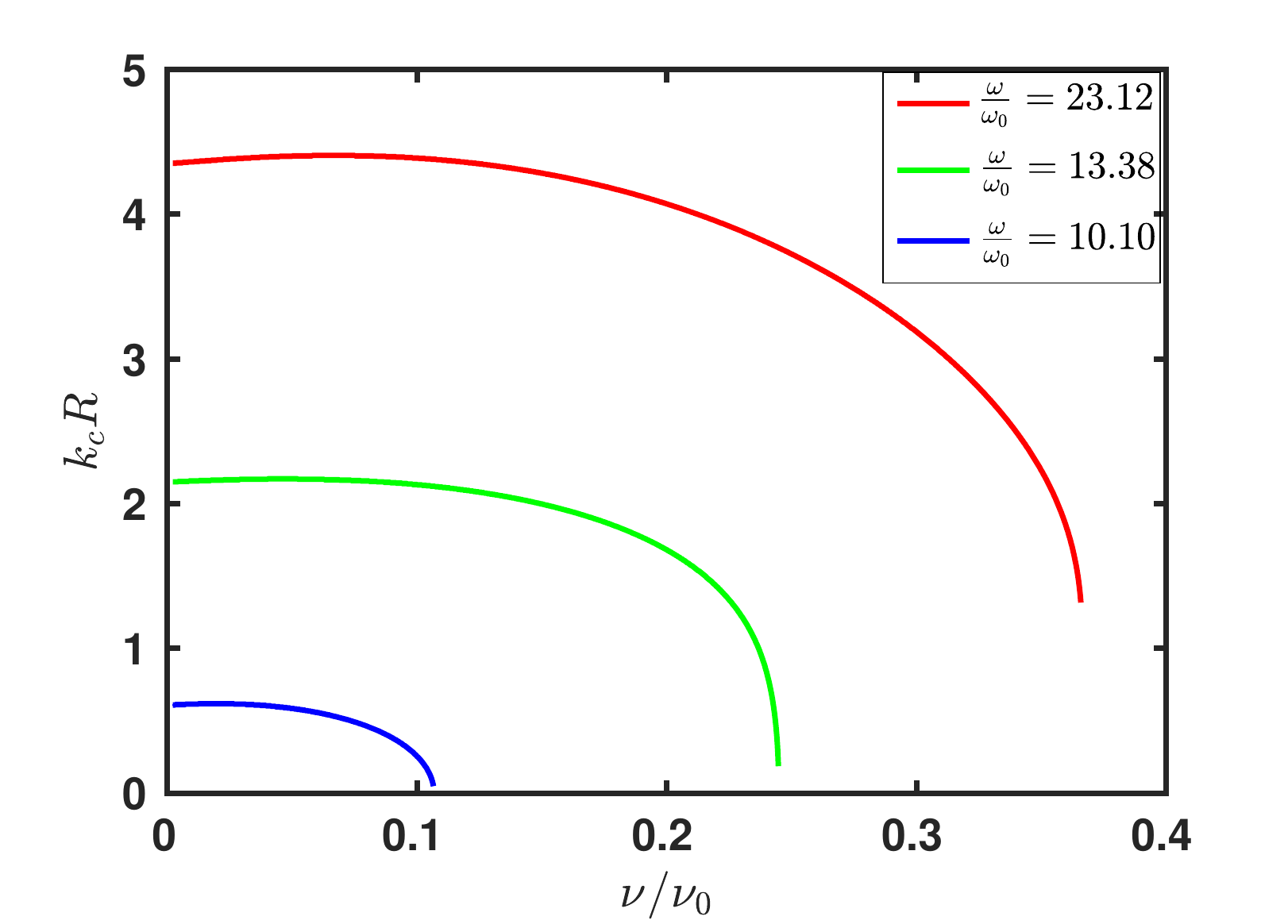}} 
		\subfigure[]{\includegraphics[width=0.48\textwidth]{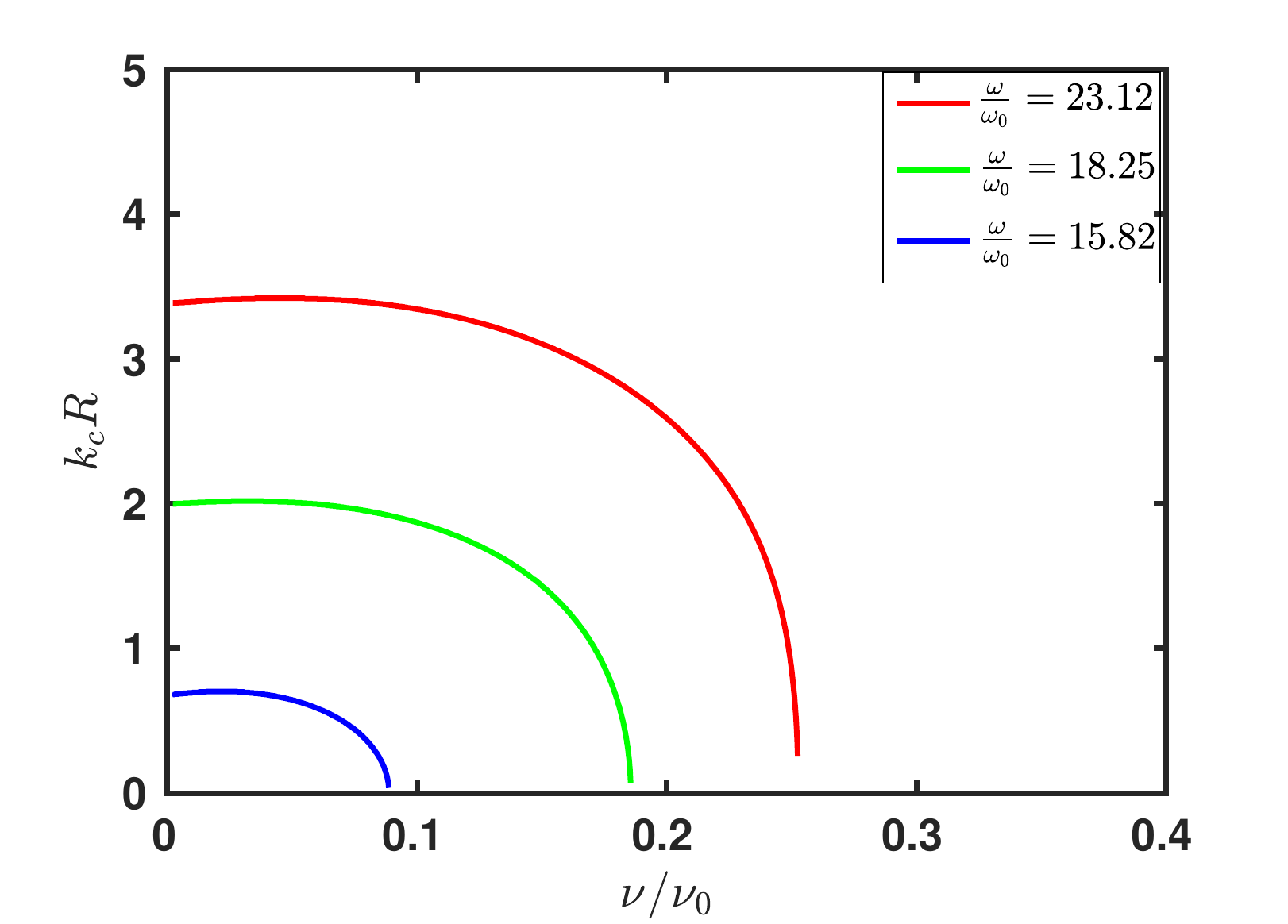}}
		\caption{(color online) Viscosity dependence of dimensionless threshold amplitude $(a_c/a_0)$  at three different forcing frequency for sub-harmonically excited standing waves pattern with (a) $m=3$ and (b) $m=4$.  Variations of the dimensionless critical axial wavenumber $(k_cR)$ with viscosity for (c) $m=3$ and (d) $m=4$.}
		\label{fig8}
	\end{figure*}
	
	The viscosity dependence of the onset parameters of the pattern with $m>1$ are showing different behavior from $m=1$. In Fig.~\ref{fig7}(a), monotonic behavior is noticed on the variation of the critical amplitude of $m=2$, but in Fig.~\ref{fig7}(b), the variation of the critical axial wavenumber is showing slightly non-monotonic behavior. 
	At lower viscosity, $k_cR$ slowly increases, and then at higher viscosity, it sharply decreases (discussed in Kumar And Tuckerman~\cite{Kumar_1994}). 
	Unlike $m=1$, there is a cut-off viscosity. After the cut-off viscosity, no minimum was found on the first subharmonic tongue. Discontinuity of the curves in Fig.~\ref{fig7} indicates the cut-off viscosity in the corresponding case.
	
	In Figs.~\ref{fig8}(a,c) the critical amplitude and the critical axial wavenumber are plotted for three different values of $\omega/\omega_0$ with viscosity for $m=3$, respectively and the same are plotted for $m=4$ in Figs.~\ref{fig8}(b,d), respectively. Figs.~\ref{fig8}(a-d) make it clear that the Onset parameters of the patterns with $m>1$ are showing a similar type of viscous dependence. Also, it is observed that for a particular forcing frequency, the cut-off viscosity decreases along with the increasing of $m$.

	\section{Summary and discussion}
	In classical RP instability, the jet is deformed to droplets. The system becomes unstable only for axis-symmetric perturbation with $kR<1$ and stable for all other perturbations. Under the influence of a time-periodic radial acceleration, the cylinder is unstable for non-axis-symmetric perturbation~\cite{Dilip,Dasgupta_2}. This is cylindrical analogous to another classical instability called  Faraday instability. In the study, Linear stability analysis is done on the viscous cylindrical fluid surface, which is subjected to a time-periodic radial acceleration. The analysis shows the viscous effect on the critical forcing amplitude and the dispersion relation of the non-axis symmetric patterns.
	
	Linear stability of the system is analyzed in the cylindrical coordinate system by considering the Floquet method of Kumar~\cite{Kumar_1996}, Kumar and Tuckerman~\cite{Kumar_1994}, Adou and Tuckerman~\cite{Adou_Tuckerman_2016} and Li et al.~\cite{Li}. Decomposing the space in cylindrical harmonics and using Floquet decomposition in time, the full linearized Navier-Stokes equations are converted to a finding of the eigenvalue problem. Solving this eigenvalue problem, the marginal stability boundaries are found for the sub-harmonic and the harmonic case. The minimum value of the first subharmonic tongue gives the onset information, i.e., the critical forcing amplitude and the critical axial wavenumber of the pattern. Pattern with the harmonic, which corresponds to a tongue where $a/a_0$ is a single-valued function of $kR$, does not excite on the cylinder. The system is unstable inside the boundary (in the parameter space ($a/a_0, kR$) it looks like a tongue) and outside it remains stable forever.
	
	Due to the effect of viscosity, the tongues are distorted, and also the onset values of the patterns change. Like viscous spherical drop~\cite{Adou_Tuckerman_2016}, viscosity has more impact on the dispersion relation of the pattern with higher values of $m$. Strong non-monotonic viscous dependence of the critical forcing amplitude of $m=1$ is observed. Most interestingly, at a certain range of viscosity where the critical amplitude of $m=1$ decreases with increasing viscosity. This type of phenomenon is new in such a problem. For $m=1$, $k_cR$ increases with viscosity sharply after initial flatness.  As usual monotonic viscous dependence is observed on the critical amplitude of the patterns with $m>1$.  At lower viscosity, $k_cR$ increases, and then at higher viscosity, it decreases sharply in every pattern with $m>1$. For a particular forcing frequency, the cut off viscosity decreases with increasing $m$.
	
	
	\begin{acknowledgements}
		{\noindent  I like to thank Krishna Kumar and Sonjoy Majumder for fruitful discussions.} 
	\end{acknowledgements}

\end{document}